\PassOptionsToPackage{prologue,svgnames}{xcolor}
\documentclass[sigconf]{acmart}

\usepackage{pgfplots}
\pgfplotsset{compat=1.6, table/search path={data/total-computation/, data/progress/, data/result-size/},}
\usepgfplotslibrary{groupplots} 
\usetikzlibrary{matrix}
\usepackage{subfig}
\usepackage{tikz}
\usepackage{multirow}
\usepackage{multicol}
\usepackage{url}
\usepackage[noend]{algpseudocode}
\usepackage{algorithmicx,algorithm}
\usepackage{amsmath}
\usepackage{makecell}

\newcommand{\Paragraph}[1]{\smallskip\noindent{\bf #1.}}
\newcommand{\eat}[1]{}

\setcopyright{rightsretained}
\acmJournal{PACMMOD}
\acmYear{2026} \acmVolume{4} \acmNumber{1 (SIGMOD)} \acmArticle{73} \acmMonth{2} \acmPrice{}\acmDOI{10.1145/3786687}

\acmConference[SIGMOD'26]{SIGMOD'26}{May 31--June 5, 2026}{Bengaluru, India}
\acmPrice{}

\begin{document}

\pagenumbering{alph}


\title{RAIRS: Optimizing Redundant Assignment and List Layout for IVF-Based ANN Search}


\author{Zehai Yang}
\orcid{0009-0007-1031-8708}
\email{yangzehai20z@ict.ac.cn}

\affiliation{%
  \institution{SKLP, ACS, Institute of Computing Technology, CAS \\\ University of Chinese Academy of Sciences}
  \streetaddress{No. 6 Ke Xue Yuan South Rd, Haidian District}
  \city{Beijing}
  \country{China}
}

\author{Shimin Chen}
\authornote{Shimin Chen is the corresponding author.}
\orcid{0009-0000-1043-6236}
\email{chensm@ict.ac.cn}

\affiliation{%
  \institution{SKLP, ACS, Institute of Computing Technology, CAS \\\ University of Chinese Academy of Sciences}
  \streetaddress{No. 6 Ke Xue Yuan South Rd, Haidian District}
  \city{Beijing}
  \country{China}
}

\begin{abstract}

%
IVF is one of the most widely used ANNS (Approximate Nearest Neighbors
Search) methods in vector databases.  The idea of redundant assignment
is to assign a data vector to more than one IVF lists for reducing the
chance of missing true neighbors in IVF search.
However, the na\"ive strategy, which selects the second IVF list based
on the distance between a data vector and the list centroids, performs
poorly.  Previous work focuses only on the inner product distance,
while there is no optimized list selection study for the most popular
Euclidean space.  Moreover, the IVF search may access the same vector
in more than one lists, resulting in redudant distance computation and
decreasing query throughput.

In this paper, we present RAIRS to address the above two challenges.
For the challenge of the list selection, we propose an optimized AIR
metric for the Euclidean space.  AIR takes not only distances but also
directions into consideration in order to support queries that are
closer to the data vector but father away from the first chosen list's
centroid.
For the challenge of redudant distance computation, we propose SEIL,
an optimized list layout that exploits shared cells to reduce repeated
distance computations for IVF search.
Our experimental results using representative real-world data sets
show that RAIRS out-performs existing redundant assignment solutions
and achieves up to 1.33x improvement over the best-performing
IVF method, IVF-PQ Fast Scan with refinement.
%

\end{abstract}



\maketitle

\setcounter{page}{1}
\pagenumbering{arabic}

\section{Introduction}
\label{sec:intro}

\eat{
Searching for nearest neighbors (NN) in high-dimensional space is widely used in real-world applications, including recommendations~\cite{col_filter_recommend}, data mining~\cite{nn_data_mining}, and information retrieval~\cite{liu2007survey}.
Recent large language model (LLM) studies incorporate external information by using K-nearest neighbors for long contexts~\cite{LLM_context} and large text corpora~\cite{LLM_corpora}.
However, the enormous size of modern datasets, coupled with the curse of dimensionality~\cite{indyk1998approximate, weber1998quantitative}, renders exact NN queries prohibitively time-consuming.
As a result, Approximate Nearest Neighbors Search (ANNS) has emerged as a feasible solution to achieve better time and accuracy trade-off.

One of the most widely adopted indices for ANNS is the inverted file (IVF) structure.  IVF-based ANNS has been optimized along four primary fronts in the literature:
quantization~\cite{VA_File,ITQ,LVQ,RabitQ,AQ,LSQ,PQ,OPQ,PolysemousCodes,NEQ,DeltaPQ,VAQ,QAQ,SparCode},
distance computation optimization~\cite{PQfs,Bolt,QuickerADC,ADSampling},
hardware acceleration~\cite{SPANN,SPFresh,FaissGPU,FPGA_ann_cvpr,danopoulos2019fpga,PQfs,Bolt,QuickerADC},
and redundant assignment~\cite{SPANN,SOAR}.
While the former three fronts have been studied in depth, the last direction sees only limited exploration.  This paper focuses on the last category.
}

\begin{figure}[t] 
  \centering 

  \subfloat[Problem of single assignment] { 
     \label{subfig:cluster}
     \includegraphics[width = 0.43 \linewidth]{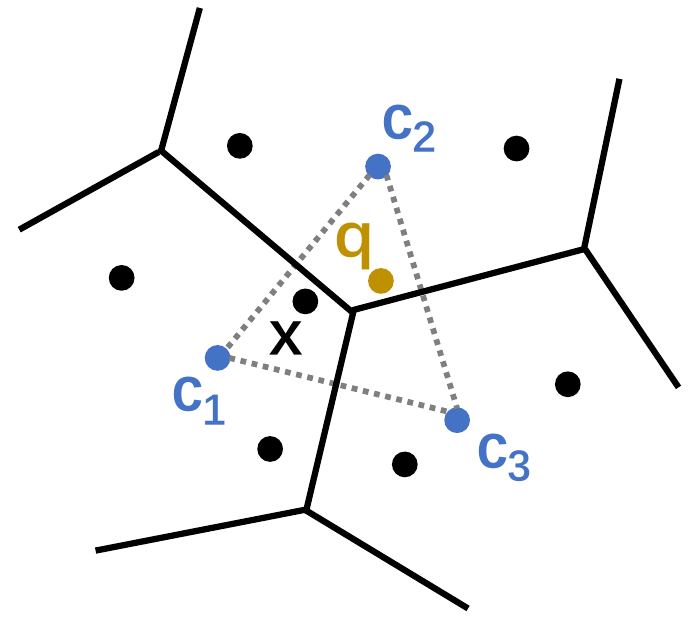}
  }
  \hfill
  \subfloat[Na\"iveRA works poorly] {
     \label{subfig:naiveRA}
     \includegraphics[width = 0.48 \linewidth]{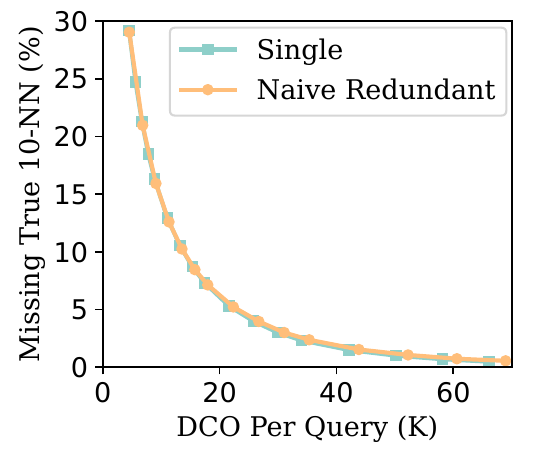}
  }
  \vspace{-0.15in} 

  \caption{Redundant assignment for IVF index.}


  \vspace{-0.25in}

\end{figure}

%
Vector search is widely used in real-world applications, including
recommendations~\cite{col_filter_recommend}, data
mining~\cite{nn_data_mining}, information
retrieval~\cite{liu2007survey}, and recently large language model
(LLM) studies~\cite{LLM_context,LLM_corpora}.  Approximate Nearest
Neighbors Search (ANNS) is the key operation in vector databases.  The
inverted file index (IVF) is one of the most widely adopted ANNS
methods.
As depicted in Figure~\ref{subfig:cluster}, during construction, IVF
computes $nlist$ clusters (a.k.a. lists) and assigns each data vector
to the list whose centroid is the closet to the vector.  For example,
$x$ is assigned to $list_1$ as $x$ lies closer to $c_1$ than any other
centroids.  Given a query $q$, IVF searches the list centroids and
chooses the top-$nprobe$ nearest lists to $q$, then traverses the
chosen lists to compute distance for all vectors in the lists.  In
this example, IVF chooses the top-2 lists (i.e., $list_2$ and
$list_3$).  Unfortunately, it fails to retrieve $x$, $q$'s true nearest
neighbor.
While increasing $nprobe$ (e.g., from 2 to 3) may help, IVF has to
traverse more lists, leading to lower query throughput.  
In this paper, we investigate an alternative solution, redundant
assignment.  The idea is to assign a data vector to more than one IVF
lists.  Suppose $x$ is assigned to both $list_1$ and $list_2$.  Then,
the traversal of $list_2$ can successfully retrieve $x$, thereby
reducing the chance of missing true neighbors.
%

\Paragraph{Challenges of Redundant Assignment}
%
%
There are two main challenges for realizing the redundant assignment
idea:

\begin{list}{\labelitemi}{\setlength{\leftmargin}{3.5mm}\setlength{\itemindent}{0mm}\setlength{\topsep}{0.5mm}\setlength{\itemsep}{0mm}\setlength{\parsep}{0.5mm}}

\item \emph{List selection strategy}:
The na\"ive strategy (Na\"iveRA) is to select the second list for a
vector based purely on its distance to the list centroids.  However,
experimental results show that Na\"iveRA can hardly improve the ANNS
performance.
Figure~\ref{subfig:naiveRA} compares Na\"iveRA with single assignment
on the SIFT1M data set for top-10 nearest-neighbor search. 
The X-axis reports the average distance computing operations (DCO) per
query, while the Y-axis shows the percentage of true neighbors missed.
We see that the two curves almost overlap, indicating that Na\"iveRA
fails to out-perform the baseline IVF single assignment.  


\item \emph{Redundant distance computation}:
With redundant assignment, IVF search may encounter the same vector in
more than one chosen lists.  For example, suppose $x$ is assigned to
$list_1$ and $list_2$.  If both lists are chosen in a query, IVF will
access $x$ twice, leading to redundant distance computation for $x$.
One fix seems to deduplicate the vector IDs from all chosen lists
before distance computation.  However, in advanced IVF methods, such
as IVF-PQ fast scan~\cite{PQfs}, lists store the vector IDs in packed blocks to
facilitate SIMD acceleration.  It is very costly to unpack the blocks
to obtain individual IDs for deduplication purposes.

\end{list}
%



\Paragraph{Our Solution: RAIRS}
To address these challenges, we propose RAIRS, consisting of the
following two optimization techniques:

%
\begin{list}{\labelitemi}{\setlength{\leftmargin}{3.5mm}\setlength{\itemindent}{0mm}\setlength{\topsep}{0.5mm}\setlength{\itemsep}{0mm}\setlength{\parsep}{0.5mm}}

\item \emph{RAIR (\underline{R}edundant list selection with
\underline{A}mplified \underline{I}nverse \underline{R}esidual)}.
We propose an optimized AIR metric for secondary list selection in the
Euclidean space.
After assigning a vector to the first list, whose centroid is the
closest, RAIR selects the second list to minimize the AIR metric,
i.e., ($||r'||^2 + \lambda r^T r'$), where $r, r'$ are the clustering
residuals of the first and the second lists, respectively, and
$\lambda$ is a constant parameter.  This metric considers not only the
distance (i.e., the first term $||r'||^2$), but also the direction
(i.e., the second term $r^T r'$).  AIR prefers a negative second term;
it selects the second list whose centroid is at an inverse direction
of the data vector compared to the first assigned list's centroid,
thereby covering queries that are closer to the vector but father away
from the first assigned list's centroid.  We formally prove the
effectiveness of AIR.
In addition, we investigate multiple list assignment by extending RAIR
to select three or more lists.



\item \emph{SEIL (\underline{S}hared-Cell \underline{E}nhanced
\underline{I}VF \underline{L}ists)}.
To alleviate redundant distance computation, we propose an optimized
list layout, SEIL.  We use $cell_{i,j}$ to denote all vectors that are
assigned to both $list_i$ and $list_j$.  Based on real-world data
analysis, we observe that there are cells that contain a large number
of vectors.  In advanced IVF methods, such as IVF-PQ fast scan, every
32 vector items are packed into a block for SIMD acceleration.  Thus,
we call cells larger than a block as large cells.  We observe that a
large fraction of vectors reside in large cells.  In light of the
observations, we design SEIL to share blocks of a large cell (e.g.,
$cell_{i,j}$) by two lists (e.g., $list_i$ and $list_j$).  Such shared
blocks enable deduplication on the blocks, thereby saving repeated
distance computation for shared blocks.
Please note that SEIL can be applied not only to RAIR, but also to any
redundant assignment strategy.

\end{list}
%

\Paragraph{Contributions}
The contributions of this paper are threefold.
First, we propose RAIR, a novel redundant assignment strategy
targeting the Euclidean space.  We formally prove the effectiveness of
the AIR assignment metric.
Second, we propose SEIL, a novel list layout to reduce repeated
distance computation caused by redundant assignment.
Finally, we perform an extensive experimental study to evaluate the
benefits of RAIRS.  Experimental results show that RAIR
out-performs existing redundant assignment strategies, and SEIL can
effectively reduce redundant distance computation.  Compared to
IVF-PQ Fast Scan with refinement (IVFPQfs), RAIRS achieves up to
1.33x speedup in query throughput while maintaining similar
recalls across the real-world data sets.

\Paragraph{Outline}
%
%
The remainder of the paper is organized as follows.
Section~\ref{sec:bg} reviews relevant background and discusses the
challenges of redundant assignment.  Section~\ref{sec:overview}
overviews the RAIRS solution, then Section~\ref{sec:rair}
and~\ref{sec:seil} propose RAIR and SEIL, respectively.  After that,
Section~\ref{sec:exp} reports the evaluation results.
Section~\ref{sec:related} discusses related work. Finally,
Section~\ref{sec:conclusion} concludes the paper.
%

\section{Background}
\label{sec:bg}

%
In this section, we review the background on ANNS, IVF, and redundant
assignment.
%

\subsection{ANN Search}
\label{subsec:bg}


\Paragraph{Problem Definition}
Given a set of $D$-dimension vectors $X=\{x_1, x_2, ..., x_n\}$, where
$x_i \in \mathbb{R}^D$ for $i=1, 2, ..., n$, and a query vector $q \in
\mathbb{R}^D$, the k Nearest Neighbors (kNN) problem finds the top-$K$
vectors that are closest to $q$.
%
%
In contemporary applications, data sets often reach massive scales
with million to billion vectors, and the vectors often consist of
hundreds to even thousands of dimensions.  The curse of
dimensionality~\cite{indyk1998approximate,weber1998quantitative} makes
it impossible to find the exact nearest neighbors without exhaustive
searching, which can be prohibitively costly.
Consequently, the focus of industry and academia has shifted towards
ANNS, sacrificing accuracy slightly for substantial
improvement in the processing speed and scalability.


\Paragraph{Distance Metrics}
The most popular distance metric is the Euclidean distance: $
dist(q,x)=\sqrt {\sum_{i=1}^D (q^{(i)}-x^{(i)})^2}, \mbox{where\ } x, q
\in \mathbb{R}^D
$.  Other common distance metrics include inner product, cosine
similarity, etc.  
In this paper, our proposed redundant selection method, RAIR, targets
the Euclidean distance, whereas the optimized list layout, SEIL, works
with all distance metrics.


\subsection{IVF (Inverted File Index)}
\label{subsec:ivf}

%
IVF is one of the most widely used ANNS methods in vector databases.
In the following, we describe the best-performing IVF variant in
practice, IVF-PQ Fast Scan with refinement~\cite{PQfs, ANN_Benchmarks}, which we use as
the baseline in our work.

\begin{list}{\labelitemi}{\setlength{\leftmargin}{3.5mm}\setlength{\itemindent}{0mm}\setlength{\topsep}{0.5mm}\setlength{\itemsep}{0mm}\setlength{\parsep}{0.5mm}}

\item \emph{Product quantization}:
PQ~\cite{PQ} is a widely adopted quantization method for accelerating
distance computation.  It divides the vector dimensions into a number
of groups (e.g., with 2 dimensions per group).  Every data vector is
divided into multiple sub-vectors accordingly.  Then, for each
dimension group, PQ partitions the sub-vectors into (e.g., 16)
clusters (e.g., using K-means).  It encodes a vector as its
sub-vectors' cluster IDs (a.k.a. code words).

\hspace{2mm}
IVF-PQ stores the PQ codes along with the vector IDs in the IVF lists.
Given a query, IVF-PQ builds LUTs (Look-Up Tables) that contain for
each dimension group the squared distance between the query sub-vector
and all cluster centroids.  To estimate $dist(q, x)$, IVF-PQ looks up
the LUT for each code word of $x$'s sub-vector and adds up the squared
distances.

\item \emph{Refinement}:
As estimated distances are not accurate, quantization methods, such as
PQ, are often combined with refinement to improve the search quality.
The idea is to retrieve a larger number of
%
(e.g., 10$\times$K for a top-$K$ query)
vectors from the quantization-enhanced IVF index, then compute the
accurate distances and re-rank the retrieved vectors to obtain the
top-$K$ results.

\item \emph{PQ Fast Scan}:
PQ Fast Scan~\cite{PQfs} is among a number of existing
techniques~\cite{PQfs,Bolt,QuickerADC} that exploit SIMD acceleration
for distance computation.  During index construction, PQ Fast Scan
organizes the items (i.e., PQ codes and vector IDs) in an IVF list
into packed fixed sized (e.g., 32-item) vector blocks to facilitate
SIMD accesses.  Then, for query processing, it loads the LUTs into the
SIMD registers, and uses SIMD instructions to compute the distances
for a block of items at a time.

\end{list}

\subsection{Redundant Assignment}
\label{subsec:bg-rs}

\begin{figure}[t]
  \centering
  \includegraphics[width=0.95 \linewidth]{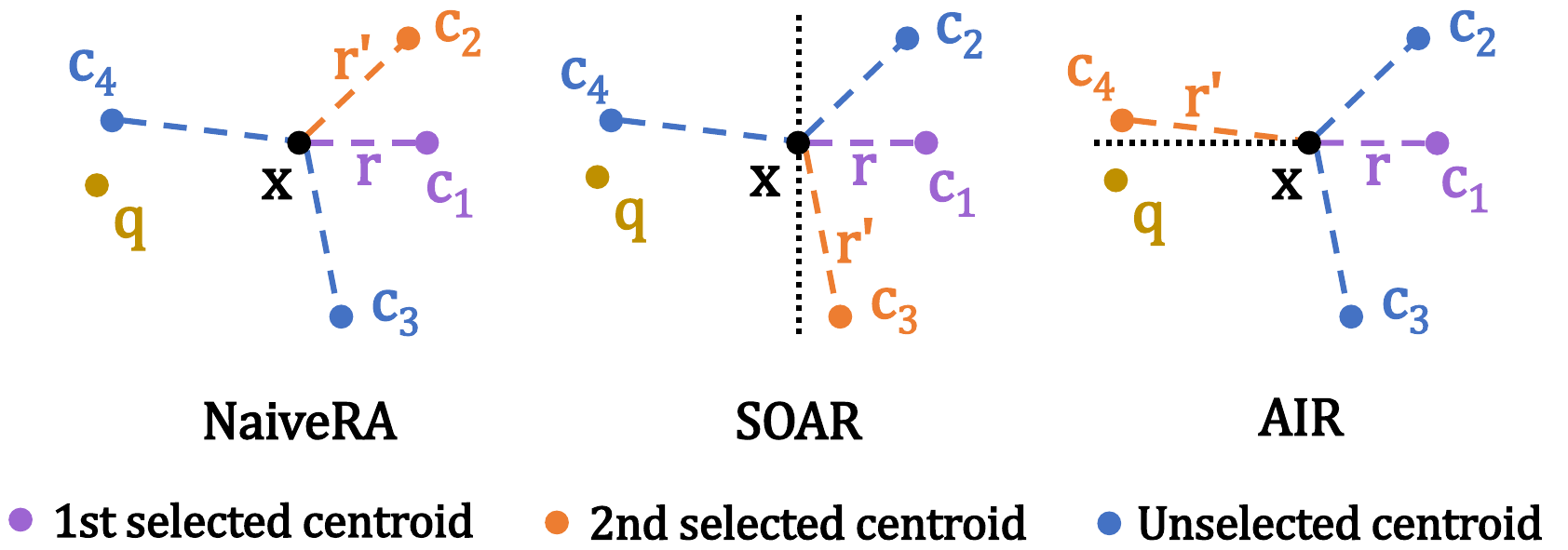}

  \vspace{-0.1in}

  \caption{Different redundant assignment strategies.}
  \label{fig:assign_strategy}

  \vspace{-0.2in}

\end{figure}

%
Redundant assignment allocates each vector item to multiple IVF lists
rather than a single list, thereby decreasing the likelihood of
overlooking true top-$K$ nearest neighbors that may originally be
assigned to only a list far from the query vector.

Na\"iveRA (Na\"ive Redundant Assignment) relies solely on the distance
for list selection.  SPANN~\cite{SPANN} uses Na\"iveRA to place a
subset of vectors in multiple lists in SSD pages.  In contrast, we
focus on memory-resident environments.  


SOAR~\cite{SOAR} is a redundant assignment strategy for the inner
product distance. Given the first list, it selects the second list
with the least $ || r' ||^2 + \lambda (\frac{r^T r'}{|| r ||})^2 $,
while $r, r'$ are the clustering residuals of the first and the second
lists, respectively.   The second term is non-negative and is
minimized when $r$ and $r'$ are close to orthogonal.

Figure~\ref{fig:assign_strategy} illustrates Na\"iveRA and SOAR.  $x$
is a vector.  $c_1$, $c_2$, $c_3$, and $c_4$ are four list centroids.
Because $c_1$ is the nearest centroid, $x$ is first assigned to
$list_1$.  However, for a query $q$, $x$ is $q$'s true nearest
neighbor, but $c_1$ is far away from $q$.  Hence, it is likely that
the IVF search for $q$ may miss $x$.  We would like to perform
redundant assignment for $x$.
As shown in Figure~\ref{fig:assign_strategy}, Na\"iveRA simply selects
the second-nearest centroid, $c_2$.  SOAR selects $c_3$, whose
residual vector $r'=x-c_3$ is close to orthogonal to the primary
residual vector $r=x-c_1$.  Unfortunately, neither $c_2$ nor $c_3$ is
ideal for $q$.

In this paper, we propose an AIR strategy optimized for the Euclidean
space.  To support queries like $q$, which are closer to the vector
but father away from the first assigned list’s centroid, AIR selects the
second list so that its residual is in the opposite direction of the
primary residual.  As shown in Figure~\ref{fig:assign_strategy}, $c_4$
is selected, which supports $q$ well.  


\section{RAIRS Overview}
\label{sec:overview}

%
We overview the data structure, the two proposed optimizations, and
the index operations of RAIRS.
%

\begin{figure}[t]
  \centering
  \includegraphics[width=0.93 \linewidth]{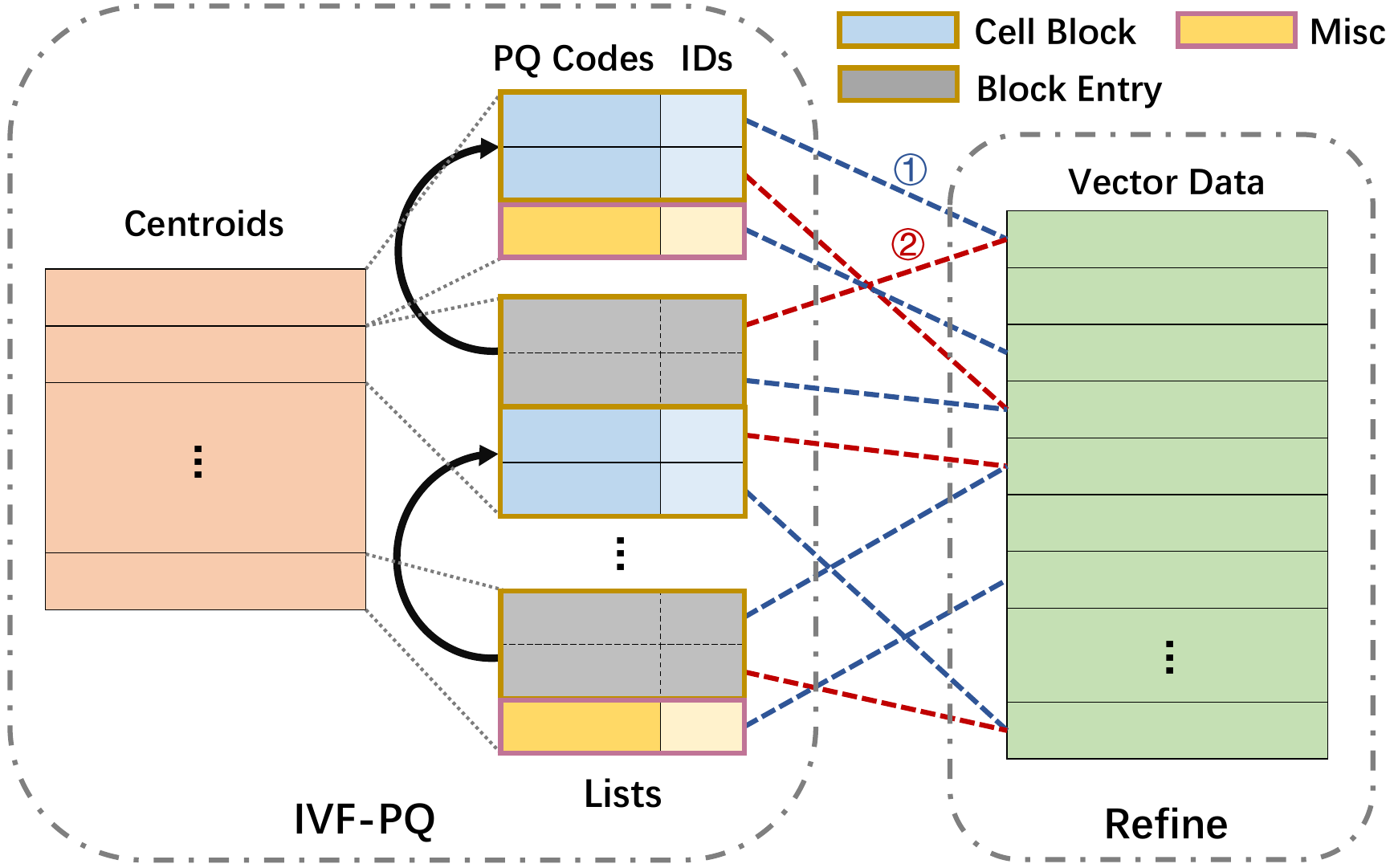}

  \vspace{-0.1in}
  \caption{Overview of the RAIRS index.}
  \label{fig:mem_layout}
  \vspace{-0.2in}

\end{figure}

\Paragraph{Data Structure}
RAIRS inherits the data structure of IVF-PQ with refinement.  As
illustrated in Figure~\ref{fig:mem_layout}, it comprises three main
components: 1) centroids, 2) inverted lists, and 3) vector data.  The
first two components form the IVF-PQ module, while the refine module
keeps the original vector data.  Each vector is assigned to up to two
lists, as depicted by the blue and red dotted lines.  The inverted
lists store PQ codes and vector IDs, consuming much less memory space
than the original vectors.  
At query time, ANNS first retrieves a set of candidate
vectors through IVF-PQ. Then, it accesses the refine module to compute
accurate distances for the candidates, and re-ranks the candidates to
obtain the final top-$K$ results. 


\Paragraph{RAIR}
Given a query $q$, IVF traverses the $nprobe$ lists whose
centroids are the closest to $q$.  However, IVF would miss $q$'s
top-$K$ nearest neighbor $x$ if $x$ is not in the closest $nprobe$
lists. To address this problem, we propose an AIR
(\underline{A}mplified \underline{I}nverse \underline{R}esidual)
metric for optimized redundant assignment in the Euclidean space.
RAIR employs the AIR metric to assign each vector to a second IVF
list.  In this way, it improves the chance of finding true top-$K$
results given the same $nprobe$.  From another angle, it can reduce
the number of traversed lists for attaining similar search accuracy,
thereby improving query throughput.  Moreover, we consider the case
where the first and the second selected lists are the same, and we
generalize RAIR to multiple assignments. (cf.
Section~\ref{sec:rair})


\Paragraph{SEIL}
We use the expression ``$x$ is in $cell_{i, j}$'' to mean that a
vector $x$ is assigned to $list_i$ and $list_j$ with redundant
assignment.  We observe that a subset of the cells contain a large
number of vectors.  There is strong skew in the number of vectors
across the cells.  
This interesting finding motivates us to optimize the list layout for
large cells in order to reduce redundant distance computation.

In the baseline, each list is divided into packed 32-item
blocks to facilitate SIMD computation by PQ Fast Scan.  A large
$cell_{i, j}$ is stored twice in both $list_i$ and $list_j$.  If the
two lists are both traversed in the same query, distance computation
will be performed twice for $cell_{i, j}$, which is wasteful.
To address this problem, we propose SEIL that stores the shared blocks
of $cell_{i, j}$ only once.  In Figure~\ref{fig:mem_layout}, the
shared cell blocks are depicted with the blue color.  The gray colored
block entry points to the shared cell block.
For a query, SEIL performs distance computation for the shared blocks
only once, decreasing redundant distance computation.
(cf. Section~\ref{sec:seil})

\Paragraph{Index Construction}
Algorithm~\ref{alg:add} shows the procedure for adding a batch of
vectors into the RAIRS index.  For each vector, the algorithm calls
\emph{RairAssign} to assign the vector to two lists (Line 4) and
computes its PQ code (Line 6).  The original vector is also appended
to the vector data to facilitate accurate distance calculations by the
refine module (Line 7).  Finally, the algorithm invokes
\emph{SeilInsert} to insert the vector items into the inverted lists
with SEIL layout optimization (Line 8).
\emph{RairAssign} and \emph{SeilInsert} will be detailed in
Section~\ref{subsec:air_ivf} and~\ref{subsec:seil_construction},
respectively.


\Paragraph{ANNS Query Processing}
Algorithm~\ref{alg:search} lists the ANNS procedure to find top-$K$
nearest neighbors for a batch of queries.  It first
computes the number of candidates ($bigK$) to retrieve from the IVF
lists as $K$ multiplied by a pre-defined $K\_FACTOR$ (e.g., 10) (Line
2).  For each query vector, the algorithm constructs the PQ distance
lookup table ($LUT$) (Line 4) and identifies $nprobe$ lists whose
centroids are closest to the query (Line 5).  Then, it calls
\emph{SeilSearch} to traverse each relevant list in the SEIL
structure, computes approximate distances based on $LUT$ and the PQ
codes, and retrieves a set of $bigK$ candidates (Line 6).  Finally, the
refine module attains the top-$K$ results based on exact distance
calculations (Line 7).  Section~\ref{subsec:seil_search} describes in
detail how \emph{SeilSearch} efficiently obtains the desired
candidates while reducing redundant distance computation.

\begin{algorithm}[t]
  \caption{Add a batch of vectors to RAIRS index.}
  \label{alg:add}
  \small
  \raggedright
  \hspace*{0.02in} \textbf{Input:} index, vecs, vec\_ids\\

  \begin{algorithmic}[1]
  \Function{AddVectors}{index, vecs, vec\_ids}
    \State assignments= []; codes = [];
    \For{(i = 0; i < vecs.len; i ++)}
      \State (listID1, listID2) = \textsc{RairAssign}(index, vecs[i]);
      \State assignments.append( \{listID1, listID2, vec\_ids[i]\} );
      \State codes.append( \textsc{PQEncoding}(vecs[i], index.code\_book) );
      \State index.vec\_data.append(vecs[i]);
    \EndFor
    \State \textsc{SeilInsert}(index, assignments, codes, vec\_ids);
    \State index.ntotal += vecs.len;
  \EndFunction
  \end{algorithmic}
\end{algorithm}

\begin{algorithm}[t]
  \caption{ANNS with RAIRS index.}
  \label{alg:search}
  \small
  \raggedright
  \hspace*{0.02in} \textbf{Input:} index, queries, K, nprobe\\
  \hspace*{0.02in} \textbf{Output:} results

  \begin{algorithmic}[1]
  \Function{RairsSearch}{index, queries, K, nprobe}
    \State bigK = K * K\_FACTOR;
    \For{(i = 0; i < queries.len; i ++)}
      \State LUT = \textsc{ComputeLookupTable}(queries[i], index.code\_book);
      \State selected\_lists = \textsc{FindNearestLists}(index, queries[i], nprobe);
      \State candidates = \textsc{SeilSearch}(index, LUT, selected\_lists, bigK);
      \State results[i] = \textsc{Refine}(index.vec\_data, queries[i], candidates, K);
    \EndFor
    \State \Return results;
  \EndFunction
  \end{algorithmic}
\end{algorithm}


\Paragraph{Applicability}
In this paper, we assume that the vector data can fit into the main
memory.  We focus on IVF-PQ Fast Scan with refinement as the baseline
to demonstrate the effectiveness of our proposed RAIR and SEIL
optimizations.

Please note that RAIR can be applied to any IVF-based indices.  The
redundant assignment strategy is orthogonal to the quantization
method, the storage medium, and the hardware optimization of the
indices.  Moreover, while SEIL is designed to support packed blocks in
PQ Fast Scan, the idea of exploiting shared cells to reduce redundant
distance computation can be generally applicable.
%
For example, for a disk-resident IVF-flat index, which
stores IVF lists on disk and centroids in memory, we can apply the
idea of SEIL by replacing the on-disk lists with shared cells of
vectors and recording the shared cell addresses along with the list
centroids in memory.


\section{Redundant List Selection with AIR}
\label{sec:rair}

We propose AIR (Amplified Inverse Residual) as an optimized list
selection metric in the Euclidean space in Section~\ref{subsec:air}.
Then, we describe and analyze the RAIR algorithm to assign a vector to
two lists in Section~\ref{subsec:air_ivf}.  Finally, we consider the
generalization of RAIR to assign a vector to multiple lists in
Section~\ref{subsec:more_lists}.

%


\subsection{Amplified Inverse Residual}
\label{subsec:air}

\begin{figure}[t]
  \centering
  \includegraphics[width= 0.9\linewidth]{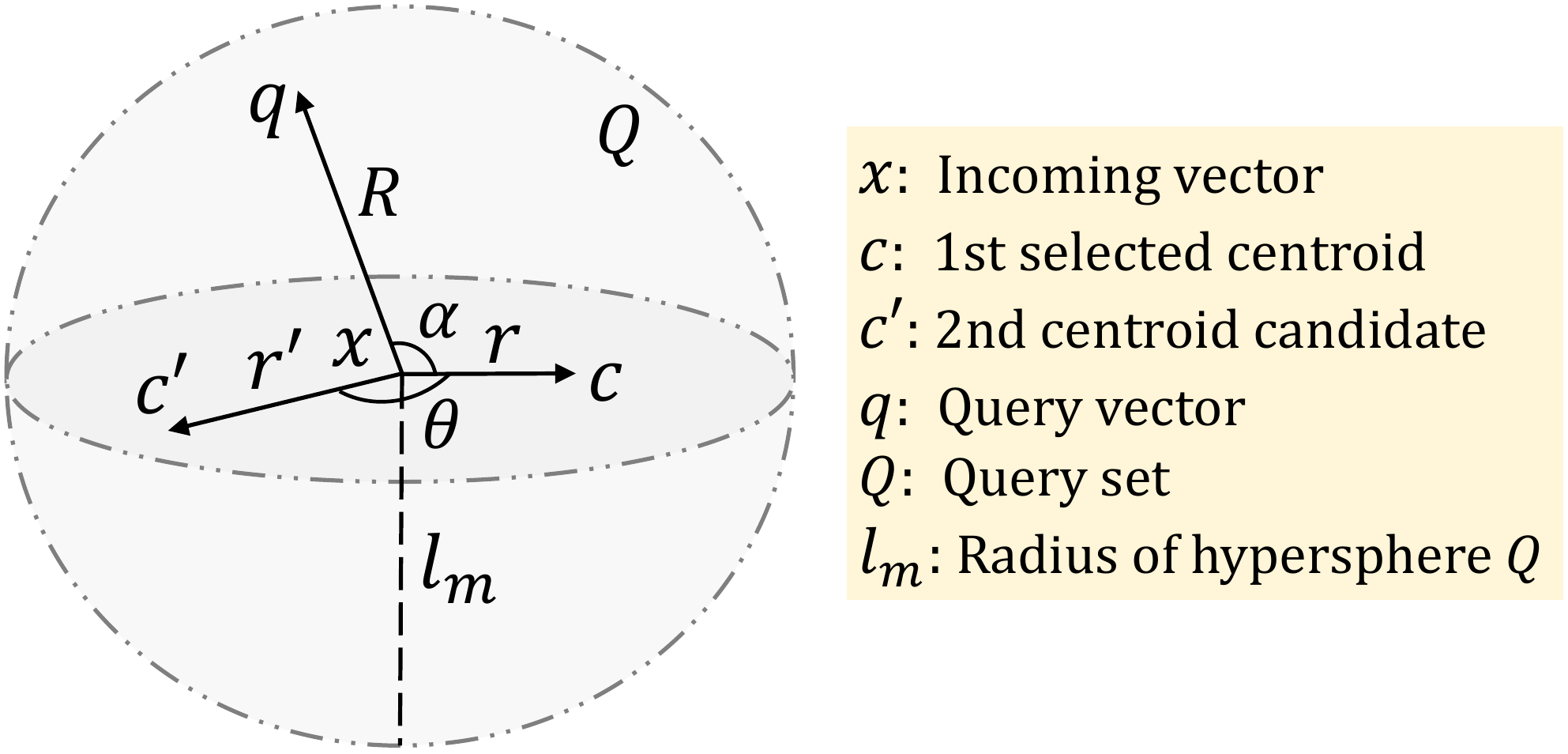}

  \vspace{-0.1in}

  \caption{Geometric relationship of vectors.}
  \label{fig:RAIR_geo}

  \vspace{-0.2in}

\end{figure}

Let $x$ be the data vector to be inserted into the IVF lists and $c$
be the centroid closest to $x$.  We consider queries within
a maximal distance $l_m$ of $x$.  Let $Q=\{q:||q-x|| \leq l_m\}$ be
the set of all queries in the hypersphere centered at $x$ with radius
$l_m$.  Suppose $q \in Q$ is a random query vector that is uniformly
distributed in $Q$. Figure~\ref{fig:RAIR_geo} depicts the geometric
relationship of the vectors.

Since $c$ is the closest centroid to $x$, the list
associated with centroid $c$ is the first selected list to assign $x$.
In most cases, representing $x$ with $c$ is satisfactory.  However,
for some query $q$, $c$ may not be an ideal representation for $x$.
That is, $c$ lies outside the nearest $nprobe$ lists of $q$, and
therefore $x$ is a true top-$K$ nearest neighbor of $q$ but does
not appear in the retrieved result with single assignment.  In such
cases, while $q$ is close to $x$, $q$ is so far away from $c$ that
there are $nprobe$ other centroids closer to $q$ than $c$.

We would like to select the second list for $x$ to accommodate such
queries that do not benefit sufficiently from the first
selected list with centroid $c$.  As shown in
Figure~\ref{fig:RAIR_geo}, let $\alpha=\angle qxc$.  $\alpha$
quantifies how \emph{unhappy} a query $q$ is with $c$ as the first
selected centroid for $x$.  This is because in the triangle $\triangle
qxc$, larger $\alpha$ indicates longer edge $qc$, which is
$dist(q,c)=||q-c||$. That is, the larger the $\alpha$, the farther
away $q$ is from $c$, and the less likely that $c$ appears in the
nearest $nprobe$ lists of $q$.

Building on this intuition, we formulate the following
loss function for selecting the second list.  Let $c'$ denote
the centroid of the second selected list.  We ensure that the second
centroid $c'$ compensates for $c$ among all queries in $Q$:
$$ L(c',c,Q) = E_{q \in Q}[{ReLU}(-\cos\angle qxc) \cdot (||q-c'||^2 -
||q-x||^2)] $$
In the loss function, the second factor, $||q-c'||^2 - ||q-x||^2$, is
easy to understand.  It expresses the preference for decreasing the
squared Euclidean distance from $q$ to $c'$ compared to $q$ to $x$.
The first factor, ${ReLU}(-\cos\alpha)$ serves as a weighting term,
indicating how important the second centroid $c'$ is to $q$.  When
$\alpha \le \frac{\pi}{2}$, it is likely that $x$ can be visited by
$q$ in the list represented by $c$.  In such cases, $-\cos\alpha \le
0$ and ${ReLU}$ returns 0.  Hence, the contribution of the second
factor is ignored, meaning that the second list is not important.  On
the other hand, if $\alpha$ exceeds $\frac{\pi}{2}$,
${ReLU}(-\cos\alpha) > 0$.  $q$ is close to $x$ but far
away from $c$. We increase the weight for such queries, giving them
higher priority during assignment.  The larger the $\alpha$, the
higher the weight.  Then, we select the second list with the least
$L(c',c,Q)$ among all lists.

We prove the following theorem to simplify the
computation of the loss function.  The full proof is provided in the
appendix.

\begin{theorem} \label{theo:rair}
For a set $Q$ of queries that are uniformly distributed in the
hypersphere centered at $x$ with radius $l_m$,
$$
L(c',c,Q) \propto ||r'||^2 + \lambda r^T r'
$$
where $r=c-x$, $r'=c'-x$, and $\lambda>0$ is a constant
factor.
\end{theorem}

\Paragraph{AIR}
Based on Theorem~\ref{theo:rair}, we set $||r'||^2 +
\lambda r^T r'$ as the metric for selecting the second list.  From the
formula, we see that the metric does not hinge solely on minimizing
$||r'||$, the distance from the second selected centroid $c'$ to $x$.
In addition, there is an added penalty term $\lambda r^T r'$.  When
$||r'||$ is fixed, having $r'$ closer to the inverse of $r$ (i.e.,
$-r$) leads to a negative second term, which reduces the loss
function.  This indicates a preference for the second residual $r'$ to
be at an inverse direction of the first residual $r$.  Hence, we call
this metric AIR (\underline{A}mplified \underline{I}nverse
\underline{R}esidual) to capture its preference for inverse residuals.

Interestingly, if we set $\lambda=0$, AIR degenerates to Na\"iveRA,
which selects the second nearest list as the second choice. In our
experiments, we set $\lambda=0.5$ by default.  We also perform an
in-depth study of the impact of $\lambda$ on ANNS performance in
Section~\ref{subsec:in_depth}.

\begin{table}[t]
  \centering
  \caption{Comparison of redundant assignment strategies.}
  \label{tab:loss}
  \vspace{-0.15in}
  \small
  \setlength{\tabcolsep}{2pt}
  \begin{tabular}{|c|c|c|c|} 
    \hline
    \textbf{Strategy} & \textbf{Na\"iveRA} & \textbf{SOAR} & \textbf{AIR} \\ 
    \hline
    \textbf{Formula} & $|| r' ||^2$ & $|| r' ||^2 + \lambda (\frac{r^T r'}{|| r ||})^2$ & $||r'||^2 + \lambda r^T r'$ \\
    \hline
    \makecell[c]{\textbf{Geometric}\\ \textbf{Interpretation}} &
    \makecell[c]{2nd nearest\\ neighbor} 
    & \makecell[c]{Prefer 2nd residual\\ orthogonal to\\ 1st residual} 
    & \makecell[c]{Prefer 2nd residual\\ opposite to\\ 1st residual} \\
    \hline
  \end{tabular}
  note: $r$ ($r'$) is the clustering residual of the first (second)
selected list.

  \vspace{-0.10in}
\end{table}

\Paragraph{Comparison of Redundant Assignment Strategies}
%
%
Table~\ref{tab:loss} compares Na\"iveRA, SOAR, and AIR.  First,
Na\"iveRA aims to minimize the distance of the list centroid for the
second selected list.  In comparison, both SOAR and AIR consider not
only distances but also the relationship between the first and the
second selected lists, as evidenced by the second term of their formula.
Second, SOAR and AIR are significantly different.  The second term in
SOAR is proportional to the squared projection of the second residual
$r'$ on the first residual $r$.  Thus, SOAR prefers $r'$ to be
orthogonal to $r$, making the second term to close to 0.  In contrast,
the second term of AIR computes the dot product of the two residuals,
and can be negative.  AIR prefers $r'$ to be at the inverse direction
of $r$.
We compare Na\"iveRA, SOAR, and AIR experimentally in
Section~\ref{sec:exp}.

\subsection{RAIR Algorithm}
\label{subsec:air_ivf}

Algorithm~\ref{alg:rair} shows the \emph{RairAssign} procedure.
Given a data vector $v$, the algorithm first obtains $N\_CANDS$ lists whose centroids are closest to $v$ (Line 2).  The $cand\_lists$ are sorted in the ascending order of the distance between a list's centroid and $v$.
Then, the AIR loss function is computed for all candidate lists (Line 3--8).
After that, the algorithm selects the primary list to assign $v$ as $cand\_lists[0]$, which is the list whose centroid is closest to $v$ (Line 9).
It determines the secondary assignment by selecting the list that minimizes the computed AIR loss function (Line 11).

\Paragraph{RAIR and Strict RAIR (SRAIR)}
In cases where the AIR loss function remains minimal for $v$’s primary
list, i.e., for any list,
$$
||r'||^2 + \lambda r^T r' \ge (1+\lambda)||r||^2,
$$
there is little benefit in assigning $v$ to any secondary list. Thus,
such vectors are stored only in their first-choice list.  This
strategy not only curtails space overhead by limiting unnecessary
redundancy but also reduces unworthy accesses to vectors when
querying, thereby potentially improving overall query performance.  We
call this strategy RAIR.  

%
In addition to RAIR, we also provide a strict version of RAIR, called
SRAIR.  SRAIR assigns a vector strictly to two lists.  That is, it
excludes the first selected list and applies AIR to select the second
list from the rest of the lists. 

The algorithm uses an $is\_strict$ flag to support both RAIR and SRAIR
(Line 10).  When $is\_strict$ is false, $start$=0 and the second list
is selected from all $N\_CANDS$ lists (Line 10--11).  When $is\_strict$
is true, $start$=1, and $cand\_lists[0]$, which is the first selected
list, is excluded from consideration (Line 10--11).
%

\Paragraph{Reducing Computation Cost with Limited Candidate Lists}
In practice, we do not compute the loss function across all $nlist$
lists for each vector.  Instead, we evaluate only the top $N\_CANDS$
nearest lists, as shown in Algorithm~\ref{alg:rair}.  Note that this
is important for large data sets, where $nlist$ is large.
\emph{FindNearestLists} can perform an ANNS rather than the exhaustive
search, thereby reducing the worst-case $O(nlist \cdot D)$ cost for
each vector.

We find that a small $N\_CANDS$ (e.g., 10) is often sufficient for the
quality of redundant assignment. For instance, in the
SIFT1M~\cite{Sift_Gist} data set, for over 99.95\% of
vectors, the minimal loss function is obtained among the top-10
nearest lists when $\lambda = 0.5$ and $nlist=1024$.  We study the
setting of $N\_CANDS$ in depth in Section~\ref{subsec:in_depth}.

\begin{algorithm}[t]
  \caption{Assign a vector to lists using RAIR.}
  \label{alg:rair}
  \small
  \raggedright
  \hspace*{0.02in} \textbf{Input:} index, v\\
  \hspace*{0.02in} \textbf{Output:} listID1, listID2\\

  \begin{algorithmic}[1]
  \Function{RairAssign}{index, v}
    \State cand\_lists = \textsc{FindNearestLists}(index, v, N\_CANDS);
    \State centroid0 = index.centroids[cand\_lists[0]];
    \State residual0 = centroid0 $-$ v;
    \For{(i = 0; i < N\_CANDS; i ++)}
      \State centroid = index.centroids[cand\_lists[i]];
      \State residual = centroid $-$ v;
      \State loss[i] = L2sqr(residual) $ +\ \lambda \cdot$ InnerProd(residual0, residual);
    \EndFor
    \State listID1 = cand\_lists[0];
    \State start = (is\_strict) ? 1 : 0;
    \State listID2 = cand\_lists[ argmin(loss[start .. N\_CANDS-1]) ];
    \State \Return listID1 < listID2 ? (listID1, listID2) : (listID2, listID1);
  \EndFunction
  \end{algorithmic}
\end{algorithm}

\Paragraph{Cost Analysis}
%
First, selecting the first list consists of calling
\emph{FindNearestLists} (Line 2) and setting $listID1$ (Line 9) in
Algorithm~\ref{alg:rair}.  Depending on the implementation, the cost
of \emph{FindNearestLists} is $O(nlist \cdot D)$ if
\emph{FindNearestLists} performs exhaustive search, or can be
decreased to $O(sublinear(nlist) \cdot D)$ if \emph{FindNearestLists}
performs ANNS.  (For example, if \emph{FindNearestLists} performs
IVF-based ANNS, the complexity can be $O(\sqrt{nlist} \cdot D)$.)

Second, the remaining Algorithm steps select the second list among
$N\_CANDS$ candidate vectors using $D$-dimensional vector
computations. Therefore, the cost is $O(N\_CANDS \cdot D)$.

Overall, the time complexity can be expressed as $O((nlist + N\_CANDS)
D)$ or $O((sublinear(nlist) + N\_CANDS) D)$ depending on the
implementation of \emph{FindNearestLists}.

Note that $N\_CANDS$ is often several orders of magnitude smaller than
$nlist$.  The additional cost of selecting the second list is often
much smaller than the cost of selecting the first list.  Thus, the cost
of list selection in redundant assignment is often close to that of
the baseline single assignment.


\subsection{Generalization to Multiple Assignments}
\label{subsec:more_lists}

\newcommand{\aggr}{\mathop{aggr}}


In the above, RAIR performs two-assignment, assigning each vector to
up to two IVF lists.  In this subsection, we generalize RAIR to
$m$-assignment, where $m \geq 3$.  

Given $(m-1)$ selected lists, we select the $m$-th centroid by
considering the losses with regard to all prior selected centroids:
\[ \begin{aligned}
&L_m(c',c_1,...,c_{m-1},Q)  = \aggr_{1\leq i \leq m-1}\ L(c',c_i,Q)
\propto ||r'||^2 + \lambda \aggr_{i}\ r_i^T r'
\end{aligned} \]
For $\aggr$, we evaluate three functions (i.e., $max$, $min$, and
$avg$) in Section~\ref{subsec:in_depth}.  Our results show that $max$
performs the best.

Note that assigning vectors to more than two lists does not
necessarily improve ANNS performance.  Distributing each vector across
additional lists can reduce the required number of traversed lists
($nprobe$) for queries. However, the average IVF list size grows,
incurring larger number of distance computing operations.  Our
experiments show that two-assignment yields the smallest number of
distance computations.


\section{Shared-Cell Enhanced IVF Lists}
\label{sec:seil}


We discuss the problems of the baseline list layout for supporting
redundant assignment in Section~\ref{subsec:naive_layout}.  Next, we
propose the SEIL layout optimization in
Section~\ref{subsec:seil_layout}.  Finally, we describe the algorithms
for the SEIL-enhanced index in Section~\ref{subsec:seil_algo}.

\begin{figure}[t]
  \centering
  \includegraphics[width=0.70 \linewidth]{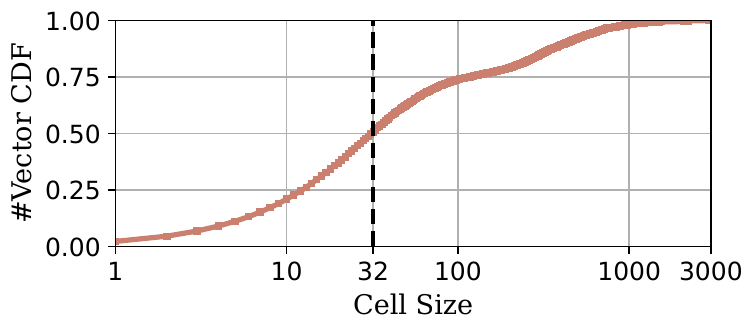}
  \vspace{-0.15in}
  \caption{\protect\mbox{Characteristics of cells after redundant assignment.}}
  \label{fig:cell_cdf}
  \vspace{-0.20in}
\end{figure}

\subsection{Challenges of Baseline List Layout}
\label{subsec:naive_layout}

Consider the case where vector $x$ is assigned to $list_i$ and
$list_j$.  In the baseline list layout, the vector item (including
$x$'s PQ code and its vector ID) is stored in both $list_i$ and
$list_j$.  PQ Fast Scan further packs every 32 vector items into a
block to facilitate SIMD acceleration in each list.  This baseline
list layout suffers from two issues: 1) redundant distance
computations at query time if both $list_i$ and $list_j$ are among the
$nprobe$ lists chosen for a query, and 2) increased space cost for
storing the vector item twice.

The first issue lowers the query throughput.  One na\"ive solution is
to collect all vector IDs from the $nprobe$ chosen lists, deduplicate
the vector IDs, then perform distance computation.  Another solution
is to build a hash table to keep track of the vector IDs whose
distances have been computed, then check the hash table on the fly to
avoid any redundant computation.  However, both solutions require
retrieving the vector IDs before distance computation.  Unfortunately,
this requires unpacking the packed blocks, which would break the SIMD
computation steps and drastically reduce the benefit of PQ Fast Scan.
Moreover, both solutions employ either sorting or hashing based
algorithms for all vector items in $nprobe$ lists, which can lead to
non-trivial additional time and space cost.

The second issue is mitigated by duplicating the vector items rather
than the $D$-dimensional vectors in the lists.  Moreover, if the first
and the second assigned lists are the same, RAIR avoids to store the
vector item twice.  Nevertheless, it would be nice to further reduce
the space cost introduced by redundant assignment.


\begin{figure}[t]
  \centering
  $\begin{array}{c}
     \includegraphics[width=\linewidth]{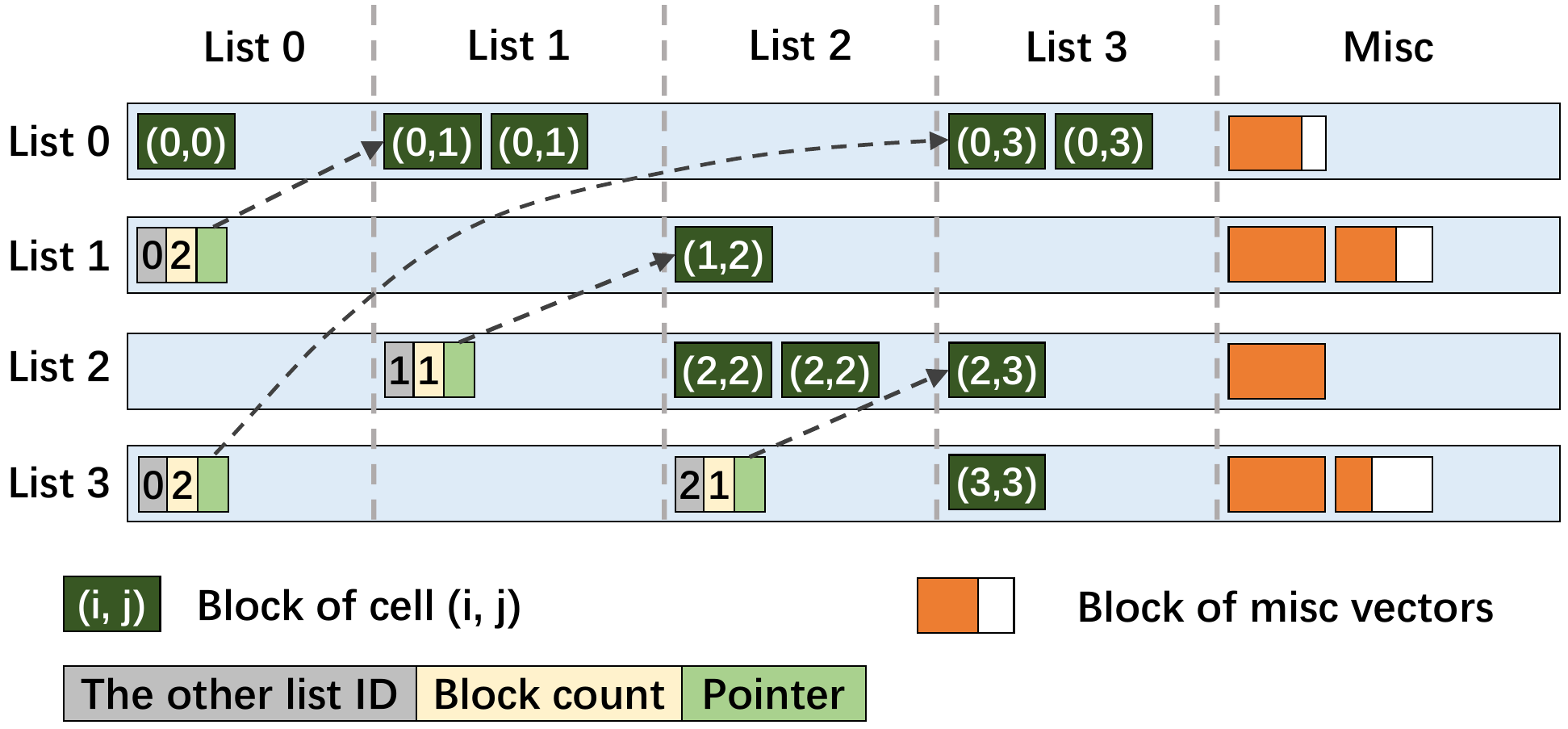}\\
     \mbox{(a) Illustration of shared cells}
     \vspace{0.05in}\\
     \includegraphics[width=\linewidth]{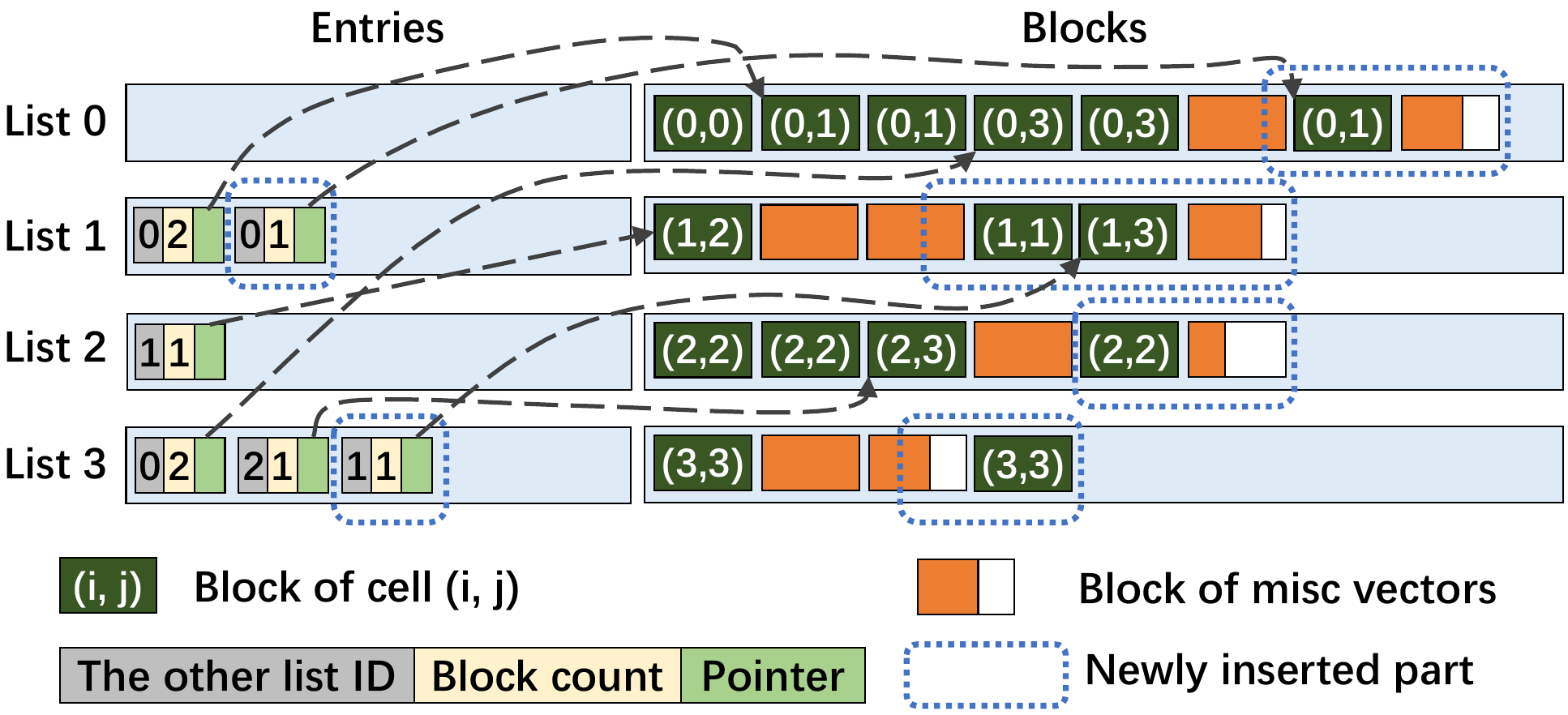}\\
     \mbox{(b) List structures after two batch of insertions}
  \end{array}$
   \vspace{-0.15in}
  \caption{SEIL list layout.}
  \label{fig:seil}
   \vspace{-0.20in}
\end{figure}

\subsection{SEIL Layout}
\label{subsec:seil_layout}

\vspace{-1mm}
%
\Paragraph{Characteristics of Cells}
We study the characteristics of the cells in
Figure~\ref{fig:cell_cdf}.  Recall that $cell_{i,j}$ contains all
vectors that are assigned to both $list_i$ and $list_j$.  Since
$cell_{i,j}$ and $cell_{j,i}$ are essentially the same, we ensure $i
\leq j$ for $cell_{i,j}$.  For a vector that is assigned to only a
single list $i$ in RAIR, we set its cell to $cell_{i,i}$.  We count
the vectors in each cell after redundant assignment.  The x-axis shows
the cell size (i.e., the number of vectors in a cell) in the
logarithmic scale.  The y-axis reports the Cumulative Distribution
Function (CDF) of how vectors are distributed across cells for the
SIFT1M data set.  1.0 corresponds to the sum of all cell sizes.

In Figure~\ref{fig:cell_cdf}, we draw a dotted line at cell size = 32.
Since a block contains 32 vector items in PQ Fast Scan, we consider a
cell to be large if its cell size $\geq$ 32.  In the figure, large
cells are to the right of the dotted line.  From the figure, we
observe that 1) large cells contain about 50\% of all vectors, and 2)
there exist very large cells that contain hundreds to thousands of
vectors.


This observation of a high degree of concentration of vectors in large
cells motivates us to exploit shared cells.  Our idea is to share the
vectors of a large $cell_{i,j}$ in both $list_i$ and $list_j$ as much
as possible for reducing redundant distance computations at query time
and decreasing the space cost due to redundant assignment.

\Paragraph{SEIL List Structure}
The SEIL list structure is depicted in Figure~\ref{fig:seil}.
Figure~\ref{fig:seil}(a) illustrates the idea of shared cells.
$cell_{i,j}$ is drawn at row $i$ column $j$.  Vectors in cells are
grouped into 32-item blocks.  Suppose a cell contains $nitems$
vectors.  Then, we generate ($nitems / 32$) blocks.  The remaining
($nitems \% 32$) vectors are appended to blocks in the miscellaneous
area.  The blocks of $cell_{i,j}$ are stored only once in memory.
They are shared by $list_i$ and $list_j$.  For example, $cell_{0,1}$'s
blocks are shared by $list_0$ and $list_1$.  The blocks are physically
stored in $list_0$.  $list_1$ contains a (the other list ID, block
count, pointer) entry that references the shared blocks in $list_0$.

%
Figure~\ref{fig:seil}(b) shows the physical list structures.  Each
list contains an array of reference entries and an array of 32-item
blocks.  Shared blocks of $cell_{i,j}$ ($i<j$) are physically stored
in $list_i$, while $list_j$ stores the reference entry/entries
pointing to the physical blocks.  The figure also depicts the
structures after two batches of insertions.  The entries and blocks of
the second batch are highlighted with dotted rounded rectangles.  They
are appended to the end of the entry arrays and block arrays.  

We would like to point out several design considerations.  
First, a reference entry can point to multiple blocks of the same
cell.  For a batch of insertions, SEIL stores multiple blocks of the
same cell contiguously.  A reference entry points to the first block
with the other list ID and pointer field.  The block count is the
number of contiguous blocks in the cell.
Second, for the new batch of vectors, a new entry referencing the same
other list can be generated.  For example, for the second batch,
$cell_{0,1}$ sees a new block.  Then, the new block is stored in
$list_0$, and a new reference entry pointing to $list_0$'s new block
is appended to $list_1$.  Now, $list_1$ contains two reference entries
both pointing to $list_0$.  
Third, the last miscellaneous block of a list may contain less than 32
vector items, which is depicted as half-filled orange rectangles.  All
other blocks are full.  For a new batch, we fill the last
miscellaneous block of the previous batch before generating new
miscellaneous blocks.
Finally, for a vector item stored in the miscellaneous blocks, we
embed its other assigned list ID in the unused high-order bits of the
vector ID.  Suppose $x$ is one of the remaining ($nitems \% 32$)
vectors in $cell_{i,j}$.  Then, we embed $j$ ($i$) when storing $x$ in
a $list_i$'s ($list_j$'s) miscellaneous block.  This simplifies the
deduplication of miscellaneous vectors.


\Paragraph{SEIL-Optimized Deduplication}
%
%
For blocks in shared cells, SEIL enables cell-level deduplication to
reduce redundant distance computation.   The basic idea is to
determine if a shared cell has been processed in the same query, and
skip the SIMD distance computation for all blocks of the shared cell
if the cell has been seen.

To achieve this, we distinguish physical blocks from reference
entries.   For physically stored blocks of shared cells, we always
perform distance computation.  For reference entries, we need to
determine whether their corresponding blocks have been accessed before
in the same query.  This is accomplished with a $listVisited$ hash
table that keeps track of the visited lists in the query.
$listVisited$ is probed with the other list ID of a reference entry.
If it exists, then the corresponding blocks have already been
processed and thus the reference entry is skipped.  Otherwise,
distance computation is carried out for the blocks represented by the
reference entry.

For vectors in the miscellaneous area, we cannot avoid distance
computation because of the packed blocks of PQ Fast Scan.  However, it
is still necessary to deduplicate the vectors after the distance
computation so that the same vector won't appear twice in the returned
result.  Instead of maintaining detailed records of all miscellaneous
vectors accessed, we exploit the above $listVisited$ hash table to
simplify the deduplication.  Basically, for each miscellaneous vector
$x$, we check whether the other list ID embedded in $x$'s vector ID
has already been accessed in $listVisited$, and immediately skip the
vector if the check returns true.


\eat{
In a redundantly assigned IVF index, a query may probe several
clusters that all contain the same vector replica.  If duplicates are
removed only after the IVFPQ phase, we must retrieve the
top-$(redundant\_factor \cdot bigK)$ instead of top-$bigK$ candidates
to guarantee that at least K unique neighbors survive deduplication.
This inflates the priority queue that tracks the current best results
and amplifies the cost of its frequent updates.  To curb this
overhead, instead, we perform deduplication before vectors enter the
intermediate result queue.  Two granularities are possible:
vector-level deduplication, which filters repeated vector ID directly
after block-level distance computation, and block-level deduplication,
which eliminates entire blocks before block-level distance
computation.
}


\subsection{SEIL Algorithms}
\label{subsec:seil_algo}

\Paragraph{Constructing SEIL-Optimized Lists}
\label{subsec:seil_construction}
Algorithm~\ref{alg:seil_insert} shows the procedure to insert a batch
of vectors into SEIL-optimized lists.  The algorithm sorts the
assignment items in ascending order so that the items in the same cell
are contiguous in the $assigns$ array (Line 8).  Then, the algorithm
scans the $assigns$ array twice in two for-loops.  

The first for-loop computes the number of shared blocks and the number
of items in the miscellaneous area for each list (Line 9--15).  Each
loop iteration processes the items in the same cell.  It obtains the
cell (Line 10), counts the number of items in the cell (Line 11),
computes the number of blocks and the remaining items (Line 12), and
updates the relevant statistics (Line 13--15).  After that, this
information is used to allocate space for the lists (Line 16).  

The second for-loop populates the lists with the items (Line 17--32).
It obtains $cell$, $nitems$, $nblocks$, and $nmisc$ in the same way as
in the first for-loop (Line 18--20), then appends shared blocks (Line
23--25) and miscellaneous items (Line 26--27) in the first list.  If
the second list is not the same as the first list, the algorithm also
appends the miscellaneous items in the second list (Line 31--32), and
creates a reference entry to point to the shared blocks of the first
list (Line 30).  Vectors within the same block have their PQ codes
arranged in the PQ Fast Scan format.

\begin{algorithm}[t]
  \caption{Insert a batch of vectors to SEIL-optimized lists.}
  \label{alg:seil_insert}
  \small
  \raggedright
  \hspace*{0.02in} \textbf{Input:} index, assigns, codes, vec\_ids\\

  \begin{algorithmic}[1]

  \Function{GetCell}{assign}
    \State \Return (assign.listID1, assign.listID2);
  \EndFunction

  \Function{CountItemsWSameCell}{assigns, i, cell}
    \For {(j = i+1; j < assigns.len; j++)}
          \State \textbf{if} (\textsc{GetCell}(assigns[j]) != cell) \textbf{then} break;
    \EndFor
    \State \Return j $-$ i;
  \EndFunction

  \Function{SeilInsert}{index, assigns, codes, vec\_ids}
    \State Sort assigns in ascending order of \{listID1, listID2, vec\_id\};
    \For {(i = 0; i < assigns.len; i += nitems)}
      \State cell = \textsc{GetCell}(assigns[i]);
      \State nitems = \textsc{CountItemsWSameCell}(assigns, i, cell);
      \State nblocks = nitems / BLK\_SZ; nmisc = nitems \% BLK\_SZ;
      \State list\_nb[cell.listID1] += nblocks; list\_nm[cell.listID1] += nmisc;
      \If {(cell.listID1 != cell.listID2)}
        \State list\_nm[cell.listID2] += nmisc;
      \EndIf
    \EndFor

    \State Allocate space according to list\_nb and list\_nm;

    \For {(i = 0; i < assigns.len; i += nitems)}
      \State cell = \textsc{GetCell}(assigns[i]);
      \State nitems = \textsc{CountItemsWSameCell}(assigns, i, cell);
      \State nblocks = nitems / BLK\_SZ; nmisc = nitems \% BLK\_SZ;

      \State list1 = index.lists[cell.listID1];
      \State bptr = list1.getNextBlockPointerInSharedCell();
      \For {(b = 0; b < nblocks; b++)}
        \State begin = i + b*BLK\_SZ; end = begin + BLK\_SZ $-$ 1;
        \State list1.appendBlockInSharedCell(assigns, begin, end);
      \EndFor

      \For {(j = i + nblocks*BLK\_SZ; j < i+nitems; j++)}
        \State list1.appendItemInMiscArea(assigns, j);
      \EndFor

      \If {(cell.listID1 != cell.listID2)}
        \State list2 = index.lists[cell.listID2];
        \State list2.appendReferenceEntry(cell.listID1, nblocks, bptr);
        \For {(j = i + nblocks*BLK\_SZ; j < i+nitems; j++)}
          \State list2.appendItemInMiscArea(assigns, j);
        \EndFor
      \EndIf

    \EndFor

  \EndFunction
  \end{algorithmic}
\end{algorithm}

\eat{
\begin{table}[t]
  \centering
  \caption{Time cost of insertion on SIFT1M.}
  \label{tab:insertion_cost}
  \vspace{-0.15in}
  \small
  \begin{tabular}{|c|c|c|} 
    \hline
    \textbf{RAIRS} & \textbf{IVFPQfs} & \textbf{HNSW} \\ 
    \hline\hline
    15.802 s & 13.969 s & 136.076 s  \\ 
    \hline
  \end{tabular}
  \vspace{-0.10in}
\end{table}
}

\Paragraph{Searching SEIL-Optimized Lists} \label{subsec:seil_search}
Algorithm~\ref{alg:seil_search} uses two structures to facilitate the
search: $rqueue$ that maintains the top-$bigK$ vectors, and
$listVisited$ that keeps track of visited lists for deduplication
purposes.  The algorithm initializes the two structures (Line 2--3),
then goes into a loop to search each selected list (Line 4--18).  

For each list, the algorithm processes the reference entries (Line
6--10), shared blocks stored in the current list (Line 11--13), and
blocks in the miscellaneous area (Line 14--17).  For the reference
entries, cell-level deduplication is used to check if all
blocks pointed to by an entry can be skipped (Line 7).  For the
physically stored shared blocks, it is guaranteed that
there is no duplication since the associated reference entry will be
checked in the other list.  For the miscellaneous area, the
vectors are deduplicated after distance computation using
$listVisited$ (Line 16).  

PQ Fast Scan is invoked for each block to efficiently compute the
distances with SIMD instructions (Line 9, 12, 15).  At the end of each
loop iteration, the current list is added to $listVisited$ (Line 18).
Finally, the algorithm returns the top-$bigK$ candidates (Line 19).

\begin{algorithm}[t]
  \caption{Searching SEIL-optimized lists.}
  \label{alg:seil_search}
  \small
  \raggedright
  \hspace*{0.02in} \textbf{Input:} index, LUTs, selected\_lists, bigK\\
  \hspace*{0.02in} \textbf{Output:} candidates\\

  \begin{algorithmic}[1]
  \Function{SeilSearch}{index, LUT, selected\_lists, bigK}
    \State rqueue.init(bigK);
    \State listVisited.init();
    \For {(each listID in selected\_lists)}
        \State list = index.lists[listID];

        /* process reference entries */
        \For {(each (otherLID, nblocks, bptr) in list.ref\_entries)}
            \If {(! listVisited.exist(otherLID))}
                \For {(b = 0; b < nblocks; b ++)}
                    \State results = \textsc{PQFastScan}(LUT, bptr[b]);
                    \State rqueue.update(results);
                \EndFor
            \EndIf
        \EndFor

        /* process shared blocks without duplication checking */
        \For {(each block in list.shared\_cell\_blocks)}
            \State results = \textsc{PQFastScan}(LUT, block);
            \State rqueue.update(results);
        \EndFor

        /* process items in miscellaneous area */
        \For {(each block in list.misc\_blocks)}
            \State results = \textsc{PQFastScan}(LUT, block);
            \State results = \textsc{RemoveVectorIfVisited}(listVisited, results);
            \State rqueue.update(results);
        \EndFor

        \State listVisited.add(listID);
    \EndFor

    \State \Return rqueue.output(bigK);

  \EndFunction
  \end{algorithmic}
\end{algorithm}

\Paragraph{Implementation: Cache Optimization for Query Batch}
%
%
ANNS can be invoked for a batch of queries, which is standard in
various ANNS benchmarks~\cite{ANN_Benchmarks, OpenAI5M}.  
Let each (query, selected list) be a computation task.   
One implementation is to group the
tasks by queries, and process all the tasks of the same query before
moving on to the next query.  However, this can be suboptimal.  In
many cases, a list is visited by multiple queries in a query batch.
Since the data accessed by a query is often much larger than the CPU
cache, the implementation incurs a lot of CPU cache misses for
accessing the same list in multiple queries.

To deal with this problem, our implementation employs an optimization
technique to improve the CPU cache performance, which is available in
Milvus~\cite{Milvus} and Faiss~\cite{Faiss}.  The idea is to group the
tasks by lists, and process all the queries for the same list back to
back.  In this way, a list stays in the CPU cache for all but the
first query searching it.  For conciseness of presentation, we have
omitted the pseudo-code of this optimization in
Algorithm~\ref{alg:seil_search}.
%

\Paragraph{Cost Analysis}
%
%
We consider the cost of $SeilInsert$ in
Algorithm~\ref{alg:seil_insert}.  Suppose $n$ vectors are to be
inserted.  For sorting $assigns$, we can employ the bucket sort with
$nlist$ buckets in two passes, which takes $O(n)$ time.  In the two
for-loops, $CountItemsWSameCell$ visits each item in the $assigns$
array, resulting in $O(n)$ cost.  Packing the $n$ vector items with
their PQ codes and vector IDs into blocks in $appendBlockInSharedCell$
and $appendItemInMiscArea$ takes $O(n \cdot D)$ cost.  The rest of the
operations are performed for each cell.  Since the number of cells is
bounded by $n$, their cost is bounded by $O(n)$.  As a result, the
cost of $SeilInsert$ is $O(n \cdot D)$, which is dominated by packing
vector items into blocks.  Our SEIL optimization avoids redundant
storage of blocks for shared cells, thereby reducing the constant
factor of this cost.

Next, we consider the cost of $AddVector$ in Algorithm~\ref{alg:add},
which performs the complete insertion procedure, involving both RAIR
and SEIL.  As described in Section~\ref{subsec:air_ivf}, the cost of
$RairAssign$ for each vector is $O((sublinear(nlist) + N\_CANDS) D)$.
For $n$ vectors, its cost is $O((sublinear(nlist) + N\_CANDS)nD)$.
The remaining operations in the for-loop of $AddVector$, including PQ
encoding for vectors, take $O(n \cdot D)$ time.  $SeilInsert$ takes
$O(n \cdot D)$ time.  Therefore, the overall cost is
$O((sublinear(nlist) + N\_CANDS)nD)$.  

On the SIFT1M data set, inserting all data vectors to the RAIRS index
takes 15.8s, while the single-assignment IVFPQfs baseline requires
14.0s and the HNSW index requires 136.0s. 
%
%
The training time is 13.3s for both RAIRS and IVFPQfs.  Therefore, the
index construction time is 29.1s and 27.3s for RAIRS and IVFPQfs,
respectively.  Although RAIRS incurs a 6.6\% slowdown relative to
IVFPQfs in index construction, it is still much faster than HNSW.
Consequently, we consider the additional construction overhead
introduced by RAIRS to be within practical bounds.
Additional results on insertions are provided in
Section~\ref{subsec:overall_perf}. 

For $SeilSearch$ in Algorithm~\ref{alg:seil_search}, let
$n_{vec\_selected}$ be the total number of vectors in the selected
lists.  Then, the worst-case cost of $SeilSearch$ is
$O(n_{vec\_selected}D)$.   Suppose $n_{vec\_shared}$ is the number of
vectors in the blocks of cells shared by two selected lists.  Then,
the SEIL-optimized cell-level deduplication reduces the cost to
$O((n_{vec\_selected} - n_{vec\_shared})D)$.


\section{Evaluation}
\label{sec:exp}

We start by describing the experimental setup in
Section~\ref{subsec:exp_setup}.  Next, we report the overall
performance of RAIRS in Section~\ref{subsec:overall_perf}, followed by
in-depth analysis of individual techniques and algorithm parameters in
Section~\ref{subsec:in_depth}.  Finally, we study the flexibility of
the SEIL optimized list layout by applying it to SOAR in
Section~\ref{subsec:soar_seil}.

\subsection{Experimental Setup}
\label{subsec:exp_setup}

\Paragraph{Machine Configuration} 
We conduct all experiments on a server equipped with an Intel(R)
Xeon(R) Platinum 8360Y CPU (2.40GHz, 36 cores, 64KB L1 cache per core,
1.25MB L2 cache per core, 54MB shared L3 cache) and 1TB 3200MT/s DDR4
memory, running CentOS 7.9.2009.  The CPU supports both
AVX2 and AVX-512 SIMD instructions\footnote{
%
%
The IndexIVFPQFastScan class in FAISS v1.8.0 uses AVX2
256-bit SIMD instructions, but the compilation of FAISS exploits
AVX-512 instructions to generate libfaiss\_avx512.so.  Our experiments
use libfaiss\_avx512.so.  In our experiments, downclocking does not
occur.}.
C/C++ code is compiled using GCC 9.3.1 with -O3.

\Paragraph{Implementation} 
We implement the RAIRS index based on Faiss v1.8.0~\cite{Faiss} and
employ OpenBLAS v0.3.3 for linear algebra support.  RAIRS is
implemented as a number of subclasses of the IndexIVF class of Faiss.
It interacts with Faiss through Faiss's public APIs.  We write
approximately 2,200 lines of code.


In the experiments, both RAIR assignment and SEIL query routines are
conducted with batches of vectors by default.  
For index construction with $AddVectors$, the set of all data vectors
are provided in a batch to build the RAIRS index, which performs RAIR
assignment and constructs SEIL-optimized lists.
For query processing, each thread processes a batch of query vectors
in parallel to maximize throughput.  We perform a bulk execution of
the $FindNearestLists$ function, retrieving the nearest lists for all
query vectors in the query batch.  Then, we employ the cache
optimization for query batches as described in
Section~\ref{subsec:seil_algo}.  We scan each list for all associated
queries, which allows the current list to remain in the CPU cache,
thereby improving overall performance.

%
To support deletion, the FAISS IVF index maintains a map from each
vector ID to the vector's list ID and in-list position.
In RAIRS, we modify the map to include up to 2 list IDs and in-list
positions for each vector ID.  
Given a vector ID to be deleted, if the in-list position is in a
shared-block, we set the corresponding ID in the packed block to an
invalid ID.  
If the position is in a misc-block, we replace this entry with the
last vector ID in misc-blocks. 
%
%
Since each vector is assigned up to two lists, RAIRS modifies up to
twice as many entries compared to the baseline IVFPQfs for a deletion.
%

\begin{table}[t]
  \centering
  \caption{Data sets used in the experiments.}
  \label{tab:datasets}
  \vspace{-0.15in}
  \small
  \setlength{\tabcolsep}{2pt}
  \begin{tabular}{|c|c|c|c|c|c|} 
    \hline
    \textbf{Data Set} & \textbf{Distance} & \textbf{\#Dim} & \textbf{\#Item} & \textbf{\#Query} & \textbf{Size} \\ 
    \hline\hline
    SIFT1M~\cite{Sift_Gist} & Euclidean & 128 & 1,000,000 & 10,000 & 488MB \\
    \hline
    SIFT1B~\cite{Sift_Gist} & Euclidean & 128 & 1,000,000,000 & 10,000 & 477GB \\
    \hline
    MSong~\cite{MSong} & Euclidean & 420 & 994,185 & 1,000 & 1.56GB \\ 
    \hline
    GIST~\cite{Sift_Gist} & Euclidean & 960 & 1,000,000 & 1,000 & 3.58GB \\
    \hline
    OpenAI~\cite{OpenAI5M} & Euclidean & 1536 & 5,000,000 & 1,000 & 28.61GB \\ 
    \hline
    T2I~\cite{T2I} & Inner product & 200 & 10,000,000 & 100,000 & 7.45GB \\ 
    \hline
  \end{tabular}
  \vspace{-0.20in}
\end{table}

\Paragraph{Solutions to Compare} 
We compare the following popular ANNS indexing methods and redundant
assignment strategies:

\begin{list}{\labelitemi}{\setlength{\leftmargin}{3.5mm}\setlength{\itemindent}{0mm}\setlength{\topsep}{0.5mm}\setlength{\itemsep}{0mm}\setlength{\parsep}{0.5mm}}

\item \textbf{IVF}~\cite{IVF}:  Class IVFFlat in Faiss 1.8.0.  This is
the plain IVF index, whose list traversal performs accurate distance
computation.  We use the same IVF parameters as RAIRS in each data
set. 

\item \textbf{HNSW}~\cite{HNSW}: Class HNSW in Faiss 1.8.0.  This is a
widely-used graph-based ANNS index.  We follow the default settings in
its original work~\cite{HNSW} (e.g., $efConstruction=500$) except for
$M$.  We set $M=32$, which is the default setting of Faiss, because it
achieves better performance than $M=16$ of the original work. 

\item \textbf{IVFPQfs}~\cite{PQfs}: The IVFPQFastScan in Faiss 1.8.0
implements IVF-PQ Fast Scan.  A refine layer is added to improve the
recalls.  The main distinctions between IVFPQfs and RAIRS are the RAIR
assignment and the SEIL layout.  IVFPQfs performs the baseline single
assignment.  We use the same parameters as RAIRS.

\item \textbf{Na\"iveRA}~\cite{SPANN}: We implement the na\"ive
strategy of redundant assignment.  It uses the IVF-PQ Fast Scan with
Refine structure and the same parameters as RAIRS. SEIL is
not enabled by default.

\item \textbf{SOARL2}~\cite{SOAR}: We replace the redundant
assignment strategy of Na\"iveRA with SOAR.  Please note that SOAR is
originally designed for the inner product distance.  Here, we
directly apply SOAR to the L2 distance in the Euclidean space.

\item \textbf{RAIR and SRAIR}: RAIR and Strict RAIR (cf.
Section~\ref{subsec:air_ivf}) without the SEIL list layout
optimization.

\item \textbf{RAIRS and SRAIRS}: RAIR and Strict RAIR with the SEIL
list layout optimization.

\end{list}


\noindent
Thread-level parallelism is handled by the Faiss library.  We ensure
that all solutions run with the same number of threads and the same
parallelization scheme.  
%
Hyper-threading is not used.
For every index with the IVF-PQ Fast Scan + Refine structure, we
consistently employ the cache optimization for query batches.

\Paragraph{Data Sets} 
We use the following representative real-world data sets covering a
diverse range of vector dimensions, vector counts, and application
scenarios in the experiments.  The features of the data sets are
summarized in Table~\ref{tab:datasets}.

\begin{list}{\labelitemi}{\setlength{\leftmargin}{3.5mm}\setlength{\itemindent}{0mm}\setlength{\topsep}{0.5mm}\setlength{\itemsep}{0mm}\setlength{\parsep}{0.5mm}}

\item \textbf{SIFT}~\cite{Sift_Gist}: SIFT (Scale-Invariant Feature
Transform) descriptors are derived from an image data set.  We use two
SIFT data sets, i.e., SIFT1M and SIFT1B, which contain 1 million and 1
billion vectors, respectively.  In SIFT1B, the original descriptors
are stored as 8-bit integers; for consistency with the other data
sets, we convert them to 32-bit floats during pre-processing.

\item \textbf{MSong}~\cite{MSong}: The million song data set contains
features for a million popular music tracks.

\item \textbf{GIST}~\cite{Sift_Gist}: GIST descriptors are generated
from an image data set, which capture the spatial structure of scenes.

\item \textbf{OpenAI}: This OpenAI embedding data set is generated
from an open sourced C4 data set from the Common Crawl data.  We
download it using the command line tool of
VectorDBBench~\cite{OpenAI5M} with L2 distance as the metric type. 

\item \textbf{T2I}~\cite{T2I}: The Yandex Text-to-Image data set
contains image embeddings as data vectors and text embeddings as
queries.



\end{list}

\noindent
Since RAIRS targets the Euclidean space, our main experiments use
SIFT, MSong, GIST, and OpenAI.  Unlike the other data sets, the
distance metric of T2I is inner product.  We use T2I to study the
applicability of SEIL to the original SOAR in
Section~\ref{subsec:soar_seil}.

\Paragraph{Parameter Setting} 
By default, we set $nlist=1024$ except for data set
OpenAI~\cite{OpenAI5M} ($nlist=2048$), T2I~\cite{T2I} ($nlist=3172$),
and SIFT1B~\cite{Sift_Gist}($nlist=32768$).  These $nlist$ values are
close to $O(\sqrt{\#Item})$ of each data set, as suggested in the
Faiss library~\cite{recommend_nlist}.
For PQ Encoding, we set the number of sub-groups
$M_{PQ}$=$\tfrac{\#Dim}{2}$, and the bit length of the code word for
each sub-group $nbits_{PQ}=4$.
In the refine module, we set $K\_FACTOR=10$ for top-1 and top-10
queries. 
%
(10 is a common setting used for faiss-ivfpqfs
experiments in ANN-Benchmark results~\cite{ANN_Benchmarks}.) 
For top-100 queries, we set $K\_FACTOR=4$ to balance the time for list
traversal and vector refinement.
For RAIRS, we set $\lambda=0.5$ and $N\_CANDS=10$ based on our
experimental study of parameter settings in
Section~\ref{subsec:in_depth}.  

At query time, we vary the search parameter (e.g., $nprobe$ in
IVF-based indices and $efSearch$ in HNSW) to achieve different
trade-offs between query speed and search quality, which contribute to
the points of the reported curves in the experimental results.

%
%
%

\begin{figure*}[t]  
  \centering
  \subfloat[Comparison with popular ANNS methods] {
    \includegraphics[width=0.3154 \linewidth]{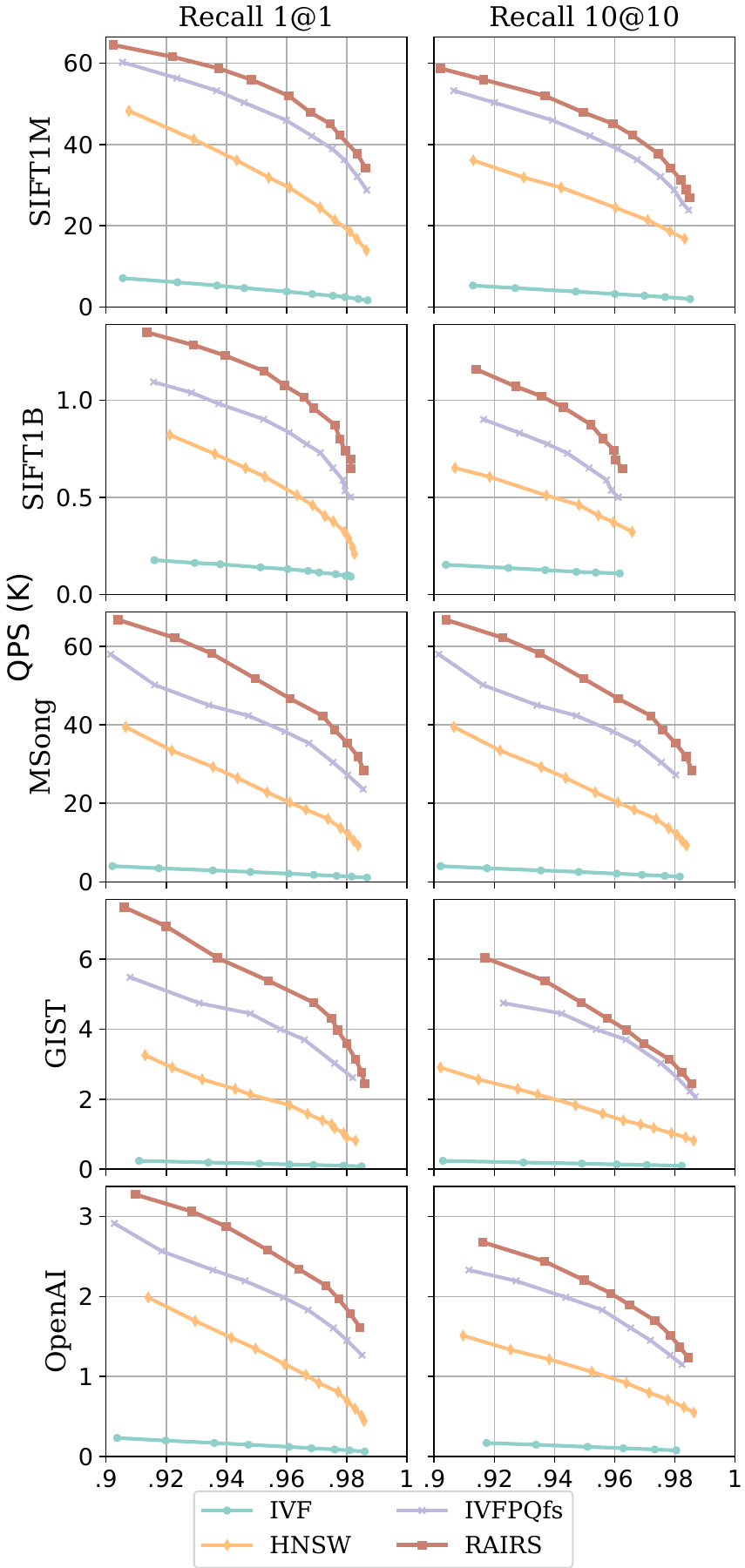}
    \label{subfig:overall_qps}
  }
  \subfloat[Varying assignment strategies] {
    \includegraphics[width=0.3403 \linewidth]{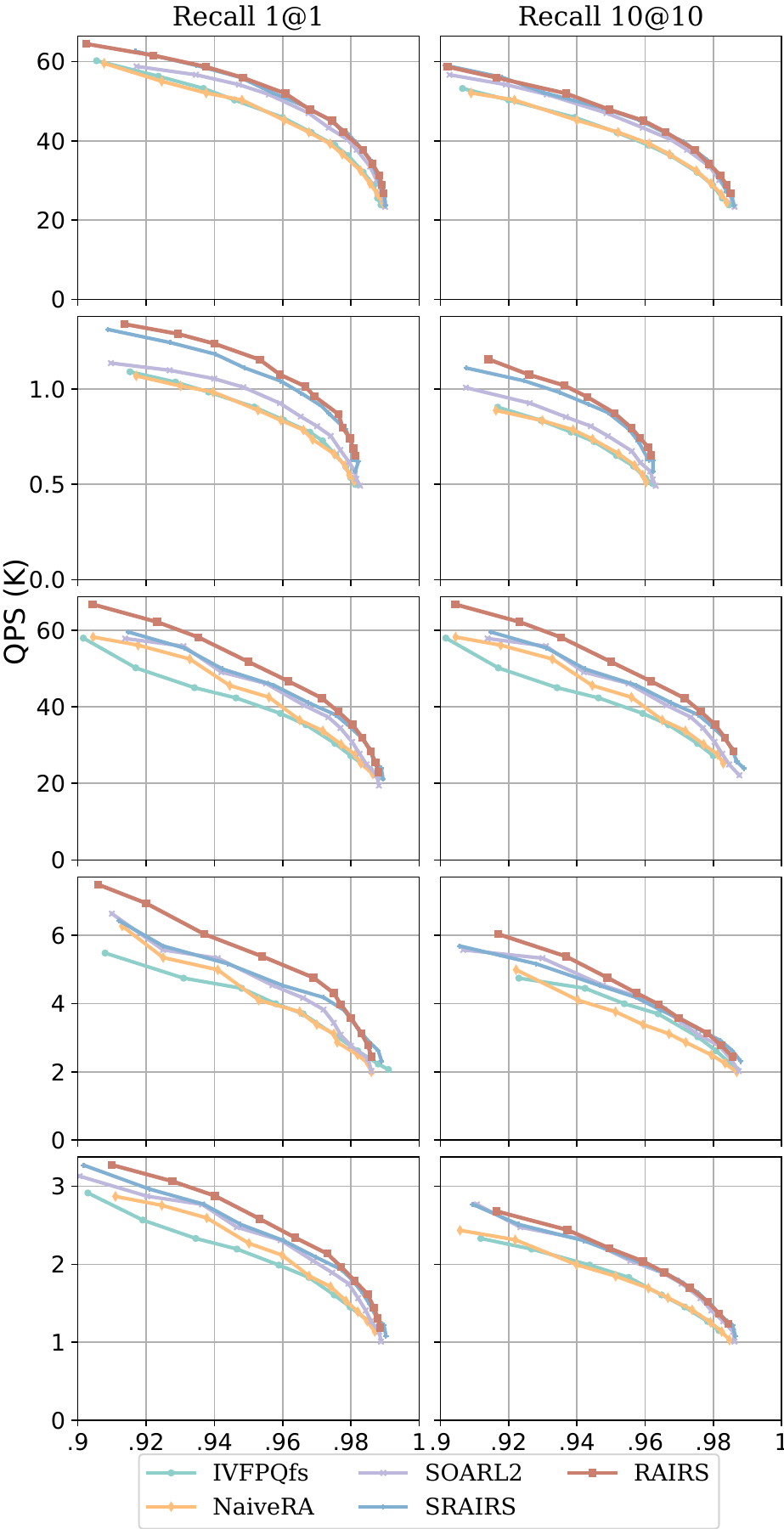}
    \label{subfig:redund_qps}
  }
  \subfloat[DCO of assignment strategies] {
    \includegraphics[width=0.3403 \linewidth]{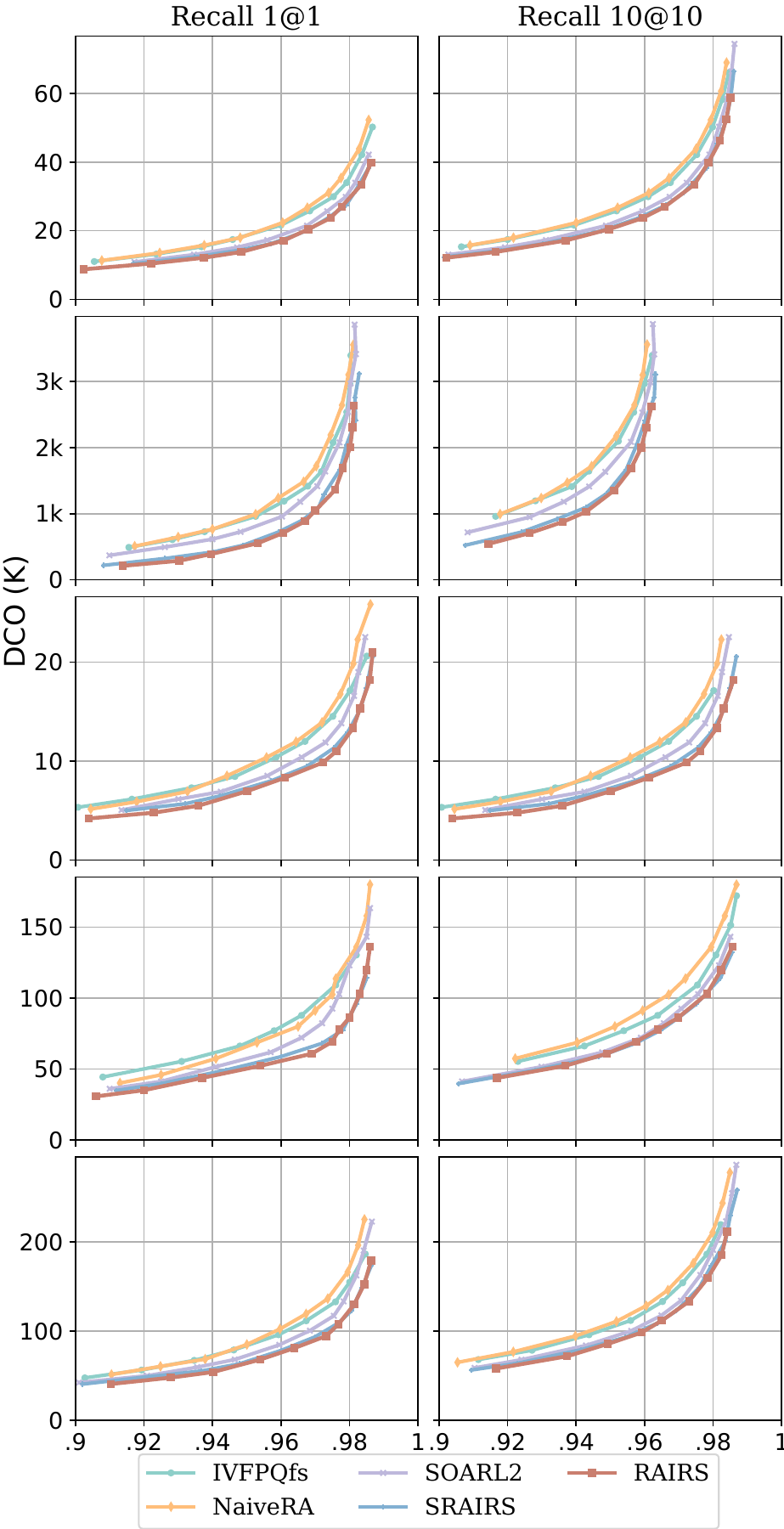}
    \label{subfig:redund_dco}
  }

  \vspace{-0.15in}

  \caption{Overall ANNS performance.}
  \label{fig:wide3}

  \vspace{-0.15in}
\end{figure*}

\begin{table}[t]
  \centering

  \caption{Percentage of vectors whose 2nd-choice centroid
under SOARL2 matches that under AIR.}

  \label{tab:assign_overlap}
  \vspace{-0.15in}
  \small
  \begin{tabular}{|c|c|c|c|c|c|} 
    \hline
    \textbf{} & \textbf{SIFT1M} & \textbf{SIFT1B} & \textbf{MSong} & \textbf{GIST} & \textbf{OpenAI} \\ 
    \hline\hline
    SOARL2 & 95.14\% & 72.10\% & 93.93\% & 91.55\% & 93.40\% \\ 
    \hline
  \end{tabular}
  \vspace{-0.10in}
\end{table}

\Paragraph{Performance Metrics} 
1) \textbf{Recall-QPS}.  Since ANNS inherently involves a trade-off
between accuracy and speed, its performance cannot be fully captured
by throughput alone. Hence, we report recall-QPS curves, which reflect
the balance between retrieval quality and processing efficiency.  For
top-$K$ queries, the recall $k$@$K$ is the average percentage of true
top-$K$ nearest neighbors in the query result.  The query throughput
is reported as QPS (Queries Per Second).
2) \textbf{Recall-DCO}.  An important cost of ANNS query processing is
distance computation~\cite{ADSampling}.   DCO is the number of
Distance Computing Operations per query.  We use DCO to understand the
effectiveness of redundant assignment strategies.  Similar to
recall-QPS curves, a recall-DCO curve is constructed by plotting the
recall $k$@$K$ against the corresponding DCO while varying the search
parameter (e.g., $nprobe$).


%
%

\subsection{Overall Performance}
\label{subsec:overall_perf}

\Paragraph{Comparison with Popular ANNS Methods}
Figure~\ref{subfig:overall_qps} shows the Recall-QPS curves of IVF,
HNSW, IVFPQfs, and RAIRS on five representative real-world data sets.
Each row of sub-figures correspond to a data set.  The left column
reports results for top-1 queries, while the right column displays
performance for top-10 queries.  In each plot, the closer to the
top-right corner, the better performance.  For each search parameter
and the resulting recall, we run the experiments for 10 times and
report the average QPS.

As shown in Figure~\ref{subfig:overall_qps}, IVFPQfs and RAIRS achieve
significantly higher performance than IVF and HNSW.  This performance
advantage comes mainly from PQ Fast Scan~\cite{PQfs}'s SIMD
acceleration, which processes packed blocks in the list traversal.
This approach is substantially faster than distance computation for
each individual vector.  In addition, the refine layer compensates for
the decrease of accuracy caused by the PQ encoding.

Among the ANNS methods, our proposed RAIRS achieves the best performance. 
It combines the RAIR redundant assignment and the SEIL optimized list layout to improve the IVF-based ANNS performance.
Compared to the second best performing method (i.e., IVFPQfs), RAIRS
achieves up to 1.33x speedup in query throughput with similar recalls
across all the real-world data sets.

\Paragraph{Comparison of Various Assignment Strategies}
Figure~\ref{subfig:redund_qps} shows the Recall-QPS curves of five
assignment strategies for top-1 and top-10 queries on the five
real-world data sets. The five solutions are all based on IVF-PQ Fast
Scan with refinement.  
From Figure~\ref{subfig:redund_qps}, we make the following
observations.
(1) Compared to the baseline single assignment (i.e., IVFPQfs),
redundant assignment with Na\"iveRA is not better, especially at high
recalls.  Na\"iveRA is actually worse than IVFPQfs for top-10 queries
on GIST.  This indicates the importance of an optimized list selection
strategy.
(2) Among all redundant assignment strategies, RAIRS achieves the best
performance across all the data sets.  
At 0.95 recall, RAIRS achieves throughput improvement of
1.07–1.33x, 1.11–1.32x, and 1.01–1.23x
compared to IVFPQfs, Na\"iveRA, and SOARL2, respectively.
%
%
%
(3) SRAIRS with strict two assignments per vector are comparable to
RAIRS on SIFT1M and SIFT1B, but worse than RAIRS on MSong, GIST, and
OpenAI.  This means that when the first and second chosen lists are
the same, it is better to store the vector item only once.
(4) RAIRS is significantly better than SOARL2 in most cases because
unlike SOAR, RAIRS is optimized for the Euclidean space.  For SIFT1M
at recall 1@1, and SIFT1M and OpenAI at recall 10@10, SOARL2 exhibits
performance comparable to RAIRS.  
AIR prefers to select a $c'$ such that $r'=c'-x$ is closer to the
inverse of $r=c-x$ (i.e., the closer $\theta$ to 180 degrees, the
better).  However, depending on the given data set and the vector in
consideration, the angle $\theta$ for the actual $c'$ may be much
smaller than 180 degrees, resulting in similar assignments as in
SOARL2.
Table~\ref{tab:assign_overlap} reports the percentage of vectors whose
2nd-choice centroid under SOARL2 matches the 2nd-choice under AIR.
We see that AIR and SOARL2 chooses the same assignment for
72.10\%--95.14\% vectors.  For the above cases where SOARL2 and RAIRS
have similar performance (i.e., SIFT1M and OpenAI), there are high
percentages of 2nd-choice matches.  From another angle, the
4.86\%--27.90\% different assignments lead to the performance benefit
of RAIRS over SOARL2.  Generally, the larger the difference, the
larger the potential benefit of RAIRS.  
%

\begin{figure}[t]
  \centering
  \includegraphics[width=0.9 \linewidth]{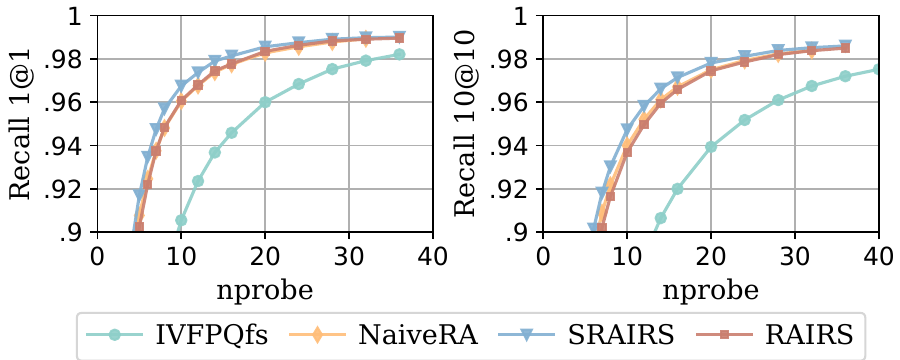}
  \vspace{-0.15in}
  \caption{Recalls varying $nprobe$ on SIFT1M.}
  \label{fig:nprobe}
  \vspace{-0.15in}
\end{figure}

\begin{figure}[t]
  \centering
  \includegraphics[width=0.9 \linewidth]{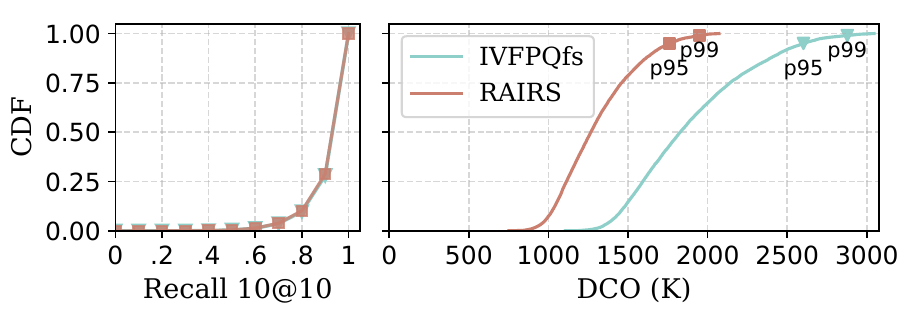}
  \vspace{-0.20in}
  \caption{Distribution of recalls and DCOs on SIFT1B.}
  \label{fig:distribution}
  \vspace{-0.15in}
\end{figure}


\Paragraph{Understanding Assignment Strategies with DCO}
Figure~\ref{subfig:redund_dco} shows the Recall-DCO curves of various
assignment strategies.  The closer to the bottom-right corner, the
better.  The Recall-DCO curves display similar trends as the
Recall-QPS curves in Figure~\ref{subfig:redund_qps}.  
Since the solutions are all based on IVF-PQ Fast Scan with refinement,
the difference in query throughput is primarily due to difference in
the number of computed distances.
At 0.95 recall, RAIRS reduces the DCO by factors of
0.64–0.83x, 0.62–0.78x, and 0.73–0.99x
compared to IVFPQfs, Na\"iveRA, and SOARL2, respectively.
%

%
Note that the benefit of RAIRS comes not only from higher recalls but
also from lower DCOs, as evidenced by Figure~\ref{subfig:redund_dco}.
RAIRS can achieve better recalls with the same DCOs, and the same
recalls with lower DCOs (which result from lower $nprobe$'s).
Figure~\ref{fig:nprobe} plots recalls while varying $nprobe$.
Comparing the $nprobe$'s for achieving similar recalls, we see that
the $nprobe$'s of SRAIRS and RAIRS are 42.3\%--46.5\% and
48.1\%--53.1\% of the setting of the baseline single assignment
IVFPQfs, respectively.
For RAIRS, since a faction of vectors are assigned only once,
achieving recalls comparable to SRAIRS can require a slightly larger
$nprobe$.
%

Given the search parameter, DCO is deterministic for a given set of
queries on a given ANNS index.  In contrast, QPS is less stable due to
run-time variations.  Therefore, for solutions based on IVF-PQ Fast
Scan with refinement, we show the ANNS performance in DCO in the rest
of the evaluation.

\Paragraph{Distribution of Recalls and DCOs}
For the SIFT1B + 0.95 recall experiment in Figure~\ref{fig:wide3}, we
compute the recall and the DCO for each of the 10,000 queries in the
top-10 search experiment.  We plot the CDF (Cumulative Distribution
Function) of recalls and DCOs in Figure~\ref{fig:distribution}.
From the figure, we see that the recall CDF curves almost overlap, but
RAIRS's DCO curve is clearly to the left of IVFPQfs's DCO curve.  This
means that compared to IVFPQfs, RAIRS significantly reduces the DCOs
for almost all queries while achieving similar recalls.
Moreover, the recall variance is low; over 89\% of queries
achieve 0.8--1.0 recalls.  The p99 DCOs of RAIRS is 1.50x of
the average.  The DCO variance across queries is moderate.
%

\begin{figure}[t]
  \centering
  \includegraphics[width=0.98 \linewidth]{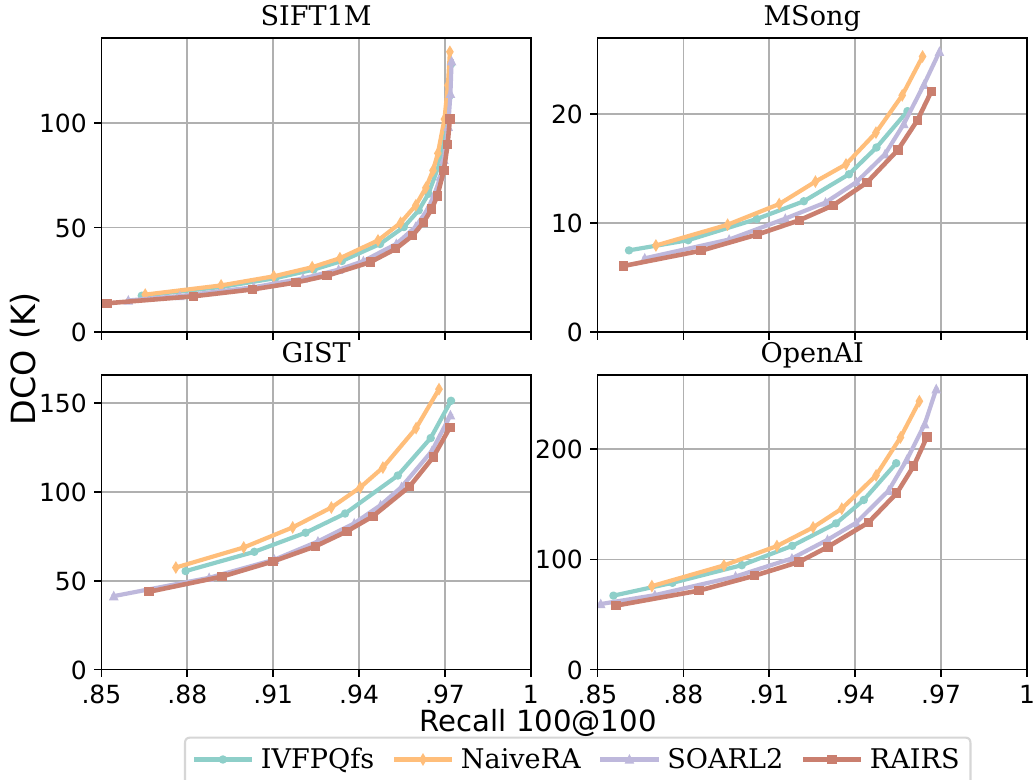}
  \vspace{-0.15in}
  \caption{Performance for top-100 queries.}
  \label{fig:top100}
  \vspace{-0.15in}
\end{figure}

\begin{figure}[t]
  \centering
  \includegraphics[width=0.9 \linewidth]{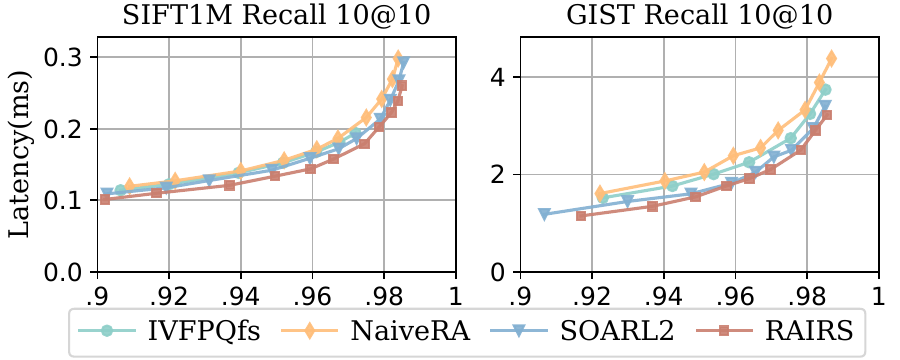}
  \vspace{-0.15in}
  \caption{Performance for one query at a time.}
  \label{fig:one_query}
  \vspace{-0.2in}
\end{figure}



\Paragraph{Performance for Top-100 Queries}
Figure~\ref{fig:top100} compares the Recall-DCO curves of IVFPQfs,
Na\"iveRA, SOARL2, and RAIRS for top-100 queries.   We see that RAIRS
achieves the best performance across all the data sets, which is
consistent with the results for top-1 and top-10 queries in
Figure~\ref{subfig:redund_dco}.


\Paragraph{Performance for One-Query-at-a-Time}
In this experiment, we run one query at a time to understand the
single-threaded query latency without the cache optimization for query
batches.  Figure~\ref{fig:one_query} shows the query latency-recall
curves for SIFT1M and GIST1M. 
We see that RAIRS achieves the lowest latency among all single
assignment and 2-assignment strategies.  The benefit of RAIRS is
similar to that seen in the batched-query scenarios.
%

\begin{figure}[t]
  \centering
  \includegraphics[width=0.9 \linewidth]{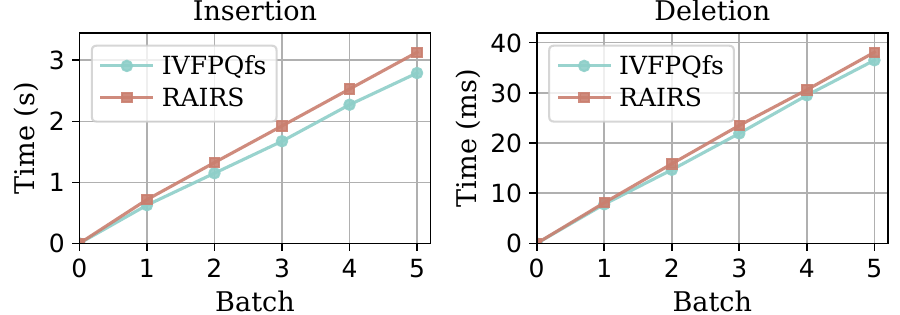}
  \vspace{-0.15in}
  \caption{Performance of insertions and deletions.}
  \label{fig:insertion_deletion}
  \vspace{-0.2in}
\end{figure}

\Paragraph{Performance of Insertions and Deletions}
Figure~\ref{fig:insertion_deletion} shows the performance of vector
insertions and deletions for RAIRS and IVFPQfs on SIFT1M.
For insertions, we construct the index with 800,000 vectors, then
perform five insertion batches, each inserting 40,000 new vectors.
For deletions, we build the index with the full data set, then execute
five batches of deletions, each removing 40,000 vectors.
Compared to the baseline IVFPQfs, RAIRS's insertion and deletion
throughput is 12.2\% and 4.4\% lower.  This is because
each vector is assigned up to two lists and RAIRS modifies up to twice
as many entries as IVFPQfs.
Please note that ANNS is typically much more frequent than vector
insertion and deletion operations.  While incurring slight extra cost
for insertions and deletions, RAIRS achieves higher ANNS query
performance.
%

\subsection{In-depth Analysis}
\label{subsec:in_depth}



\begin{figure}[t]
  \centering
  \subfloat[ANNS performance] {
    \includegraphics[width=0.60 \linewidth]{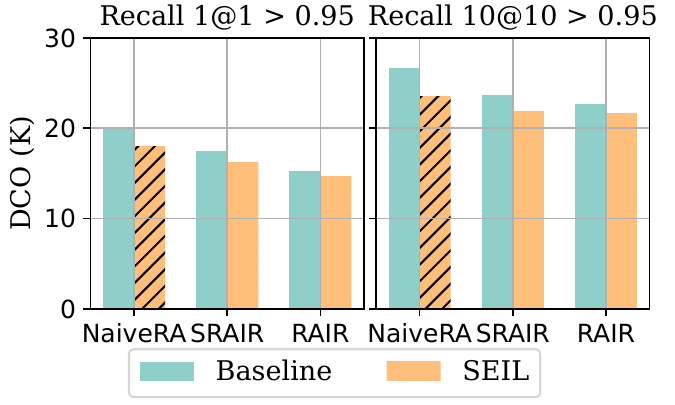}
    \label{fig:seil_ablation}
  }
  \subfloat[Memory cost] {
    \includegraphics[width=0.36 \linewidth]{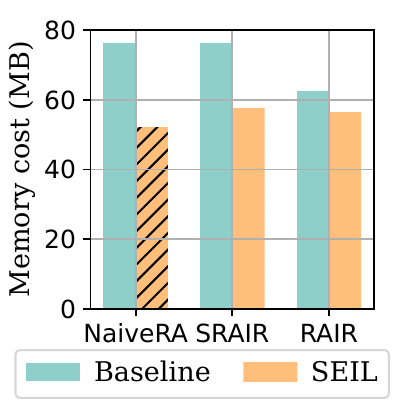}
    \label{fig:seil_ablation_mem}
  }

  \vspace{-0.1in}

  \caption{Ablation study for RAIR and SEIL on SIFT1M.}
  \label{fig:ablation}

  \vspace{-0.1in}
\end{figure}

%

\begin{table}[t]
  \centering
  \caption{Memory cost for data sets with Euclidean distance.}
  \label{tab:overall_mem}
  \vspace{-0.15in}
  \small
  \setlength{\tabcolsep}{2pt}
  \begin{tabular}{|c|c|c|c|c|c|} 
    \hline
    \textbf{Data Set} & \textbf{IVFPQfs} & \textbf{Na\"iveRA} &
\textbf{Na\"iveRA+SEIL} & \textbf{RAIR} & \textbf{RAIRS} \\ 
    \hline\hline
    \textbf{SIFT1M}   & 38.2 MB          & 76.4 MB          &     52.0 MB      &   62.5 MB     & 56.25 MB \\ 
    \hline
    \textbf{SIFT1B}   & 37.3 GB          & 74.6 GB          &       42.9 GB       &   69.7 GB     & 43.3 GB \\
    \hline
    \textbf{MSong}    & 107.2 MB         & 214.5 MB         &    145.0 MB     &  162.8 MB     & 152.4 MB \\
    \hline
    \textbf{GIST}     & 240.6 MB         &  478.5 MB         &     325.0 MB     &  403.3 MB     & 358.7 MB \\
    \hline
    \textbf{OpenAI}   & 1.83 GB          & 3.66 GB          &    2.39 GB       &   3.15 GB     &  2.53 GB \\
    \hline
  \end{tabular}
  \vspace{-0.10in}
\end{table}

\Paragraph{Ablation Study for RAIR and SEIL}
Figure~\ref{fig:seil_ablation} compares the DCO of Na\"iveRA, SRAIR,
and RAIR with and without the SEIL list layout optimization on the
SIFT1M data set.  For each compared solution, we report the DCO for
the setting where the query recall just exceeds 95\%.  In the figures,
the lower the DCO, the better. 

We see that RAIR out-performs Na\"iveRA with or without SEIL.  Without
SEIL, RAIR cuts down DCO by 24.0\% and 14.9\% compared to Na\"iveRA
for top-1 and top-10 queries, respectively.  With SEIL, the
improvement is by 18.8\% and 8.0\%, respectively.

Moreover, SEIL effectively reduces redundant distance computation.
For top-1 and top-10 queries, SEIL reduces DCO by 4.1\%--12.0\% for
Na\"iveRA, SRAIR, and RAIR.


Finally, comparing RAIR and SRAIR, we see that RAIR achieves lower
DCO.  When the first and the second assigned lists are the same, RAIR
performs single assignment, while SRAIR strictly selects a second list
from the rest of the lists to assign the vector.  The results in
Figure~\ref{fig:seil_ablation} indicates that this strict strategy is
sub-optimal.

\begin{figure}[t]
  \centering
  \includegraphics[width=0.93 \linewidth]{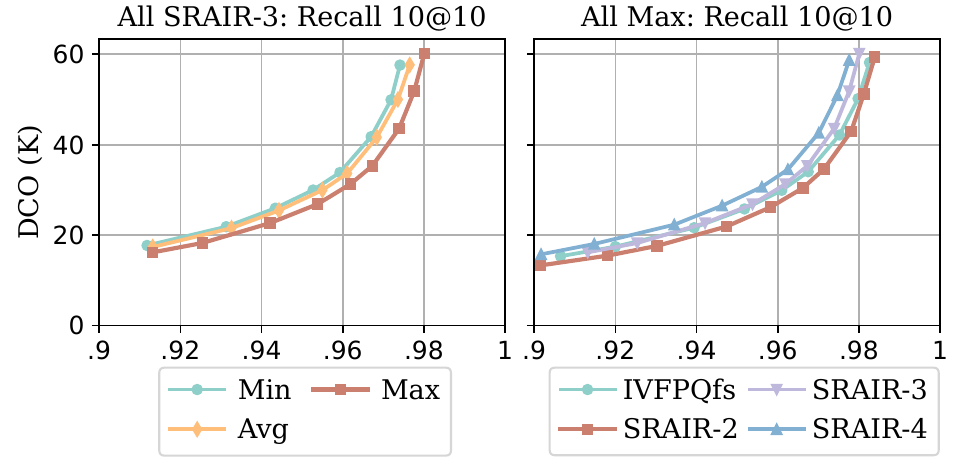}
  \vspace{-0.2in}
  \caption{Multiple assignment on SIFT1M.}
  \label{fig:more_assignment}
  \vspace{-0.25in}
\end{figure}

\begin{figure}[t]
  \centering
  \subfloat[Recall-DCO curve varying $\lambda$] {
  \includegraphics[width=0.9 \linewidth]{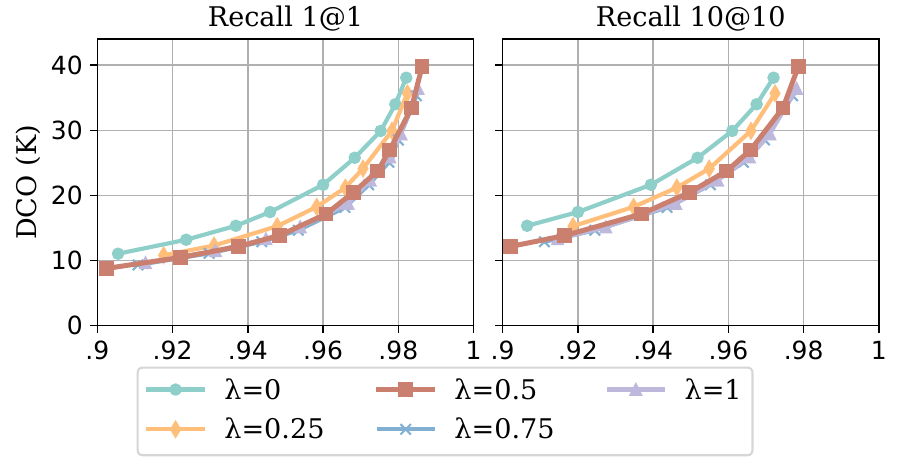}
  \label{fig:lambda}
  }
  \vspace{-0.15in}

  \subfloat[Distribution of rank of list minimizing AIR] {
  \includegraphics[width=0.875 \linewidth]{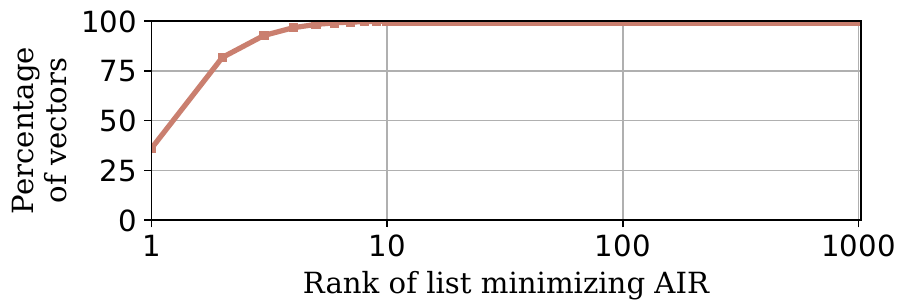}
  \label{fig:ncands}
  }

  \vspace{-0.15in}

  \caption{Parameter study (SIFT1M).}
  \label{fig:param}

  \vspace{-0.15in}
\end{figure}

\Paragraph{Memory Cost}
Figure~\ref{fig:seil_ablation_mem} displays the memory cost of
Na\"iveRA, SRAIR, and RAIR with and without SEIL on SIFT1M.
Table~\ref{tab:overall_mem} reports the memory cost of the baseline
IVPQfs and four redundant assignment solutions for five representative
data sets.  Since the refine layer is the same, we exclude the refine
layer and report only the space of the IVF-PQ module.

We see that Na\"iveRA doubles the memory space used in the baseline
IVPQfs due to redundant assignment.  SEIL stores shared blocks of
large cells only once, thereby significantly reducing the memory cost
of redundant assignment.  Overall, SEIL reduces the memory cost by
6.4\%--42.5\% for Na\"iveRA, SRAIR, and RAIR.

Note that the refine layer stores all the vector data.  Compared to
the refine layer, the baseline IVF-PQ module takes 6.4\%--7.5\% space,
and the additional space for RAIRS is 1.2\%--3.7\%.  Thus, RAIR trades
off slight extra space for significantly better ANNS performance.


\Paragraph{Multiple Assignment}
The left part of Figure~\ref{fig:more_assignment} compares the three
aggregation functions (i.e., $max$, $min$, and $avg$) for 3-assignment
on SIFT1M.  $max$ achieves the lowest DCO.
The right part of Figure~\ref{fig:more_assignment} compares single
assignment (i.e., IVFPQfs), 2-assignment, 3-assignment, and
4-assignment.  We choose the $max$ aggregation function.   SEIL is
disabled since it is designed for 2-assignment.  We consider the
strict strategy (i.e., SRAIR) because RAIR essentially blurs the
distinction among multiple assignments.  From the figure, we see that
SRAIR-2 is the best performing.  This result indicates that over two
assignment is unnecessary.

\Paragraph{Parameter Study for $\lambda$}
Figure~\ref{fig:lambda} shows recall-DCO curves of RAIRS on SIFT1M
while varying $\lambda$ from 0 to 1.  As $\lambda$ increases, the
curve shifts to the bottom-right, showing improved performance.  The
curve stops improving when $\lambda$ reaches 0.5.  Thus, we set the
default value of $\lambda$ to 0.5.


%

\Paragraph{Parameter Study for $N\_CANDS$}
We modify Algorithm~\ref{alg:rair} so that $RairAssign$ retrieves all
the lists as the candidates in the ascending order of the distance
between the list centroid and the vector to insert, and computes the
list that minimizes the AIR metric.  This gives the true rank in the
sorted lists that should be assigned by AIR.
Figure~\ref{fig:ncands} depicts the Cumulative Distribution Function
(CDF) of the true rank for all vectors in SIFT1M.  We see that in
99.95\% of the cases, the true rank $\leq$ 10.  This means that
considering the top-10 nearest lists achieves the correct assignment
for most cases.  Hence, we set $N\_CANDS=10$ by default.

\begin{figure}[t]
  \centering
  \includegraphics[width=0.9 \linewidth]{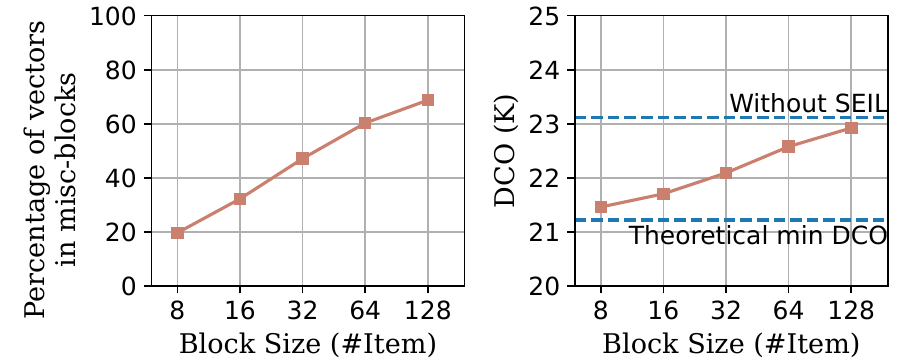}
  \vspace{-0.16in}
  \caption{Varying block size. (Default block size = 32)}
  \label{fig:block_size}
  \vspace{-0.15in}
\end{figure}

\Paragraph{Parameter Study for PQ Block Size}
%
%
To understand the impact of block size on SEIL, we vary the block
size, study the change of the number of vectors in misc-blocks, and
compute the resulting DCOs for a given $nprobe$.
As shown in Figure~\ref{fig:block_size}, as the block size increases,
there are more vectors in misc-blocks.  This is because the number of
large cells (whose size $\geq$ block size) decreases.  SEIL performs
more redundant DCOs for the vectors in misc-blocks.
%


\begin{figure}[t]
  \centering
  \includegraphics[width=0.9 \linewidth]{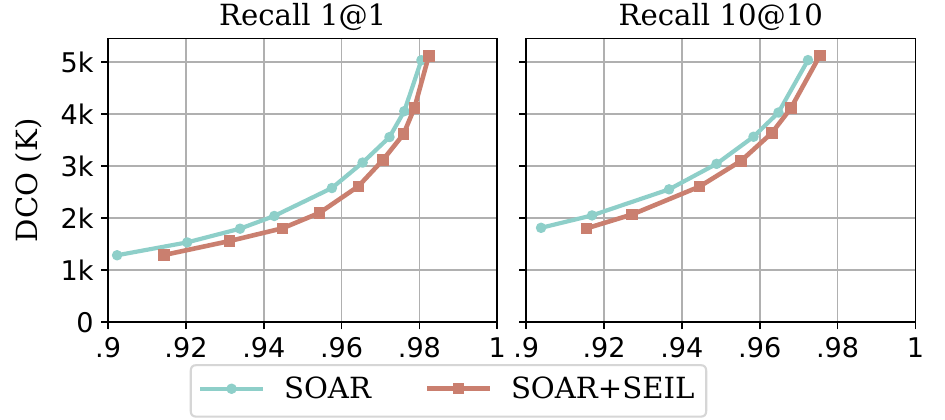}
  \vspace{-0.18in}
  \caption{Applying SEIL to SOAR on T2I.}
  \label{fig:soar_seil}
  \vspace{-0.15in}
\end{figure}

\subsection{Applying SEIL to SOAR}
\label{subsec:soar_seil}

We apply SEIL to SOAR, which targets the inner product distance.
Figure~\ref{fig:soar_seil} compares the recall-DCO curves of SOAR and
SOAR+SEIL on T2I, which uses the inner product distance as the
distance metric.  

As shown in Figure~\ref{fig:soar_seil}, we see that SEIL significantly
reduces the DCO of SOAR.  This result indicates the applicability of
the SEIL list layout optimization to various redundant assignment
strategies and distance metrics.

\section{Related Work}
\label{sec:related}

\Paragraph{ANNS Methods}
ANNS support has been implemented in both vector databases and general-purpose database systems~\cite{Milvus,Milvus_v2,PASE,AnalyticDB-V,vearch,VBASE,Starling}.
Existing ANNS methods are divided into four categories: tree-based~\cite{KDTree,BallTree,VPTree,HieKMeans,FLANN}, hashing-based~\cite{LSH,LSH_forest,Multi_probe_LSH,L2H_survey,SpectralHashing,PCAH,GQR_L2H}, graph-based~\cite{HNSW,RNG,FANNG,tau_MNG,NSG,DiskANN,HM-ANN,Song}, and cluster-based~\cite{IVF,IMI,Bliss,ScaNN} approaches.
IVF-based ANNS methods~\cite{IVF}, which are a representative type of cluster-based approaches, are among the most widely used and highest performing ANNS methods.  In this paper, we mainly focus on IVF-based ANNS.

\Paragraph{Optimization Approaches for IVF-Based ANNS Methods}
There are four main approaches to optimizing IVF-based ANNS: 1)
quantization, 2) distance computation optimization, 3) hardware
acceleration, and 4) redundant assignment.  We review related work on
approach 1--3 in the following.  Redundant assignment is the focus of
this work. We have discussed and experimentally compared the existing
redundant assignment strategies.


\emph{Approach 1: Quantization}.
Quantization techniques are widely used to mitigate storage overhead and accelerate similarity search.  Existing quantization methods can be categorized by the granularity of dimensional grouping:
single-dimension quantization (including VA\_File~\cite{VA_File}, ITQ~\cite{ITQ}, LVQ~\cite{LVQ}, and RabitQ~\cite{RabitQ}),
global quantization (including AQ~\cite{AQ} and LSQ~\cite{LSQ}),
and product quantization (including PQ~\cite{PQ}, OPQ~\cite{OPQ}, PolysemousCodes~\cite{PolysemousCodes}, NEQ~\cite{NEQ}, DeltaPQ~\cite{DeltaPQ}, VAQ~\cite{VAQ}, QAQ~\cite{QAQ}, and SparCode~\cite{SparCode}).

\emph{Approach 2: Sampling-Based Distance Computation}.
Given the high dimensionality, distance computation often dominates the ANNS query response time~\cite{ADSampling}.  ADSampling~\cite{ADSampling} leverages the Johnson–Lindenstrauss (JL) lemma and enable early termination by sampling a subset of dimensions during distance comparisons, effectively reducing the number of computed dimensions.

\emph{Approach 3: Hardware Acceleration}.
Various hardware features have been studied in the context of IVF-based ANNS.
SPANN~\cite{SPANN,SPFresh} exploits SSDs to reduce the in-memory footprint and provide data persistence.
GPUs~\cite{FaissGPU} and FPGAs~\cite{FPGA_ann_cvpr, danopoulos2019fpga} have been shown to significantly accelerate the ANNS processing.
PQ Fast Scan~\cite{PQfs}, Bolt~\cite{Bolt}, and Quicker ADC~\cite{QuickerADC} exploit SIMD instructions to accelerate the distance computation. 

\emph{Relationship to Our Work:}
In general, our proposed ideas (i.e., RAIR and SEIL) are complementary
to the above three optimization approaches.  In this paper, we
specifically use IVF with PQ quantization, refinement, and PQ fast
scan as the baseline, which is one of the best performing IVF-based
methods in practice.  We have tailored our SEIL optimization to
support PQ fast scan.


\section{Conclusion}
\label{sec:conclusion}

In conclusion, we propose RAIRS, combining two optimization techniques
(i.e., RAIR and SEIL), for IVF-based ANNS in this paper.  RAIR
exploits redundant assignment to improve performance for queries that
are poorly served by the first assigned lists.  We formally derive the
AIR list selection metric.  Moreover, SEIL optimizes the list layout
and exploits shared cells to reduce redundant distance computation.
Our experimental results on representative real-world data sets
confirm that RAIR and SEIL can effectively reduce DCO and improve
ANNS performance for IVF-based ANN search.

%
%


\bibliographystyle{ACM-Reference-Format}
\bibliography{ref}

@String{Computing = "Computing" }

@String{Computer = "{IEEE} Computer" }

@String{Springer = "Springer-Verlag" }

@inproceedings{LLM_corpora,
  title={Improving language models by retrieving from trillions of tokens},
  author={Borgeaud, Sebastian and Mensch, Arthur and Hoffmann, Jordan and Cai, Trevor and Rutherford, Eliza and Millican, Katie and Van Den Driessche, George Bm and Lespiau, Jean-Baptiste and Damoc, Bogdan and Clark, Aidan and others},
  booktitle={International conference on machine learning},
  pages={2206--2240},
  year={2022},
  organization={PMLR}
}

@article{LLM_context,
  title={Memorizing transformers},
  author={Wu, Yuhuai and Rabe, Markus N and Hutchins, DeLesley and Szegedy, Christian},
  journal={arXiv preprint arXiv:2203.08913},
  year={2022}
}

@article{PQ,
  title={Product quantization for nearest neighbor search},
  author={Jegou, Herve and Douze, Matthijs and Schmid, Cordelia},
  journal={IEEE transactions on pattern analysis and machine intelligence},
  volume={33},
  number={1},
  pages={117--128},
  year={2010},
  publisher={IEEE}
}

@inproceedings{OPQ,
  title={Optimized Product Quantization for Approximate Nearest Neighbor Search},
  author={Ge, Tiezheng and He, Kaiming and Ke, Qifa and Sun, Jian},
  booktitle={Proceedings of the 2013 IEEE Conference on Computer Vision and Pattern Recognition},
  pages={2946--2953},
  year={2013}
}

@inproceedings{VA_File,
  title={A quantitative analysis and performance study for similarity-search methods in high-dimensional spaces},
  author={Weber, Roger and Schek, Hans-J{\"o}rg and Blott, Stephen},
  booktitle={VLDB},
  volume={98},
  pages={194--205},
  year={1998}
}

@inproceedings{AQ,
  title={Additive quantization for extreme vector compression},
  author={Babenko, Artem and Lempitsky, Victor},
  booktitle={Proceedings of the IEEE Conference on Computer Vision and Pattern Recognition},
  pages={931--938},
  year={2014}
}

@inproceedings{LSQ,
  title={Revisiting additive quantization},
  author={Martinez, Julieta and Clement, Joris and Hoos, Holger H and Little, James J},
  booktitle={Computer Vision--ECCV 2016: 14th European Conference, Amsterdam, The Netherlands, October 11-14, 2016, Proceedings, Part II 14},
  pages={137--153},
  year={2016},
  organization={Springer}
}

@inproceedings{PolysemousCodes,
  title={Polysemous codes},
  author={Douze, Matthijs and J{\'e}gou, Herv{\'e} and Perronnin, Florent},
  booktitle={Computer Vision--ECCV 2016: 14th European Conference, Amsterdam, The Netherlands, October 11-14, 2016, Proceedings, Part II 14},
  pages={785--801},
  year={2016},
  organization={Springer}
}

@article{DeltaPQ,
  title={DeltaPQ: lossless product quantization code compression for high dimensional similarity search},
  author={Wang, Runhui and Deng, Dong},
  journal={Proceedings of the VLDB Endowment},
  volume={13},
  number={13},
  pages={3603--3616},
  year={2020},
  publisher={VLDB Endowment}
}

@article{ITQ,
  title={Iterative quantization: A procrustean approach to learning binary codes for large-scale image retrieval},
  author={Gong, Yunchao and Lazebnik, Svetlana and Gordo, Albert and Perronnin, Florent},
  journal={IEEE transactions on pattern analysis and machine intelligence},
  volume={35},
  number={12},
  pages={2916--2929},
  year={2012},
  publisher={IEEE}
}

@article{RabitQ,
  title={RaBitQ: Quantizing High-Dimensional Vectors with a Theoretical Error Bound for Approximate Nearest Neighbor Search},
  author={Gao, Jianyang and Long, Cheng},
  journal={Proceedings of the ACM on Management of Data},
  volume={2},
  number={3},
  pages={1--27},
  year={2024},
  publisher={ACM New York, NY, USA}
}

@inproceedings{NEQ,
  title={Norm-explicit quantization: Improving vector quantization for maximum inner product search},
  author={Dai, Xinyan and Yan, Xiao and Ng, Kelvin KW and Liu, Jiu and Cheng, James},
  booktitle={Proceedings of the AAAI Conference on Artificial Intelligence},
  volume={34},
  number={01},
  pages={51--58},
  year={2020}
}

@inproceedings{QAQ,
  title={Query-aware quantization for maximum inner product search},
  author={Zhang, Jin and Lian, Defu and Zhang, Haodi and Wang, Baoyun and Chen, Enhong},
  booktitle={Proceedings of the AAAI Conference on Artificial Intelligence},
  volume={37},
  number={4},
  pages={4875--4883},
  year={2023}
}

@inproceedings{VAQ,
  title={Fast adaptive similarity search through variance-aware quantization},
  author={Paparrizos, John and Edian, Ikraduya and Liu, Chunwei and Elmore, Aaron J and Franklin, Michael J},
  booktitle={2022 IEEE 38th International Conference on Data Engineering (ICDE)},
  pages={2969--2983},
  year={2022},
  organization={IEEE}
}

@article{LVQ,
  title={Similarity Search in the Blink of an Eye with Compressed Indices},
  author={Aguerrebere, Cecilia and Bhati, Ishwar Singh and Hildebrand, Mark and Tepper, Mariano and Willke, Theodore},
  journal={Proceedings of the VLDB Endowment},
  volume={16},
  number={11},
  pages={3433--3446},
  year={2023},
  publisher={VLDB Endowment}
}

@inproceedings{SparCode,
  title={Beyond Two-Tower Matching: Learning Sparse Retrievable Cross-Interactions for Recommendation},
  author={Su, Liangcai and Yan, Fan and Zhu, Jieming and Xiao, Xi and Duan, Haoyi and Zhao, Zhou and Dong, Zhenhua and Tang, Ruiming},
  booktitle={Proceedings of the 46th International ACM SIGIR Conference on Research and Development in Information Retrieval},
  pages={548--557},
  year={2023}
}

@article{KDTree,
  title={Multidimensional binary search trees used for associative searching},
  author={Bentley, Jon Louis},
  journal={Communications of the ACM},
  volume={18},
  number={9},
  pages={509--517},
  year={1975},
  publisher={ACM New York, NY, USA}
}

@inproceedings{VPTree,
  title={Data structures and algorithms for nearest neighbor search in general metric spaces},
  author={Yianilos, Peter N},
  booktitle={Proceedings of the fourth annual ACM-SIAM symposium on Discrete algorithms},
  pages={311--321},
  year={1993}
}

@inproceedings{BallTree,
  title={Efficient clustering and matching for object class recognition.},
  author={Leibe, Bastian and Mikolajczyk, Krystian and Schiele, Bernt},
  booktitle={BMVC},
  pages={789--798},
  year={2006}
}

@inproceedings{HieKMeans,
  title={Fast approximate nearest neighbors with automatic algorithm configuration.},
  author={Muja, Marius and Lowe, David G},
  booktitle={International Conference on Computer Vision Theory and Applications},
  volume={2},
  number={331-340},
  pages={2},
  year={2009}
}

@article{FLANN,
  title={Scalable nearest neighbor algorithms for high dimensional data},
  author={Muja, Marius and Lowe, David G},
  journal={IEEE transactions on pattern analysis and machine intelligence},
  volume={36},
  number={11},
  pages={2227--2240},
  year={2014},
  publisher={IEEE}
}

@article{LSH,
  title={Near-optimal hashing algorithms for approximate nearest neighbor in high dimensions},
  author={Andoni, Alexandr and Indyk, Piotr},
  journal={Communications of the ACM},
  volume={51},
  number={1},
  pages={117--122},
  year={2008},
  publisher={ACM New York, NY, USA}
}

@inproceedings{Multi_probe_LSH,
  title={Multi-probe LSH: efficient indexing for high-dimensional similarity search},
  author={Lv, Qin and Josephson, William and Wang, Zhe and Charikar, Moses and Li, Kai},
  booktitle={Proceedings of the 33rd international conference on Very large data bases},
  pages={950--961},
  year={2007}
}

@inproceedings{LSH_forest,
  title={LSH forest: self-tuning indexes for similarity search},
  author={Bawa, Mayank and Condie, Tyson and Ganesan, Prasanna},
  booktitle={Proceedings of the 14th international conference on World Wide Web},
  pages={651--660},
  year={2005}
}

@article{SpectralHashing,
  title={Spectral hashing},
  author={Weiss, Yair and Torralba, Antonio and Fergus, Rob},
  journal={Advances in neural information processing systems},
  volume={21},
  year={2008}
}

@inproceedings{PCAH,
  title={Annosearch: Image auto-annotation by search},
  author={Wang, Xin-Jing and Zhang, Lei and Jing, Feng and Ma, Wei-Ying},
  booktitle={2006 IEEE Computer Society Conference on Computer Vision and Pattern Recognition (CVPR'06)},
  volume={2},
  pages={1483--1490},
  year={2006},
  organization={IEEE}
}

@article{L2H_survey,
  title={A survey on learning to hash},
  author={Wang, Jingdong and Zhang, Ting and Sebe, Nicu and Shen, Heng Tao and others},
  journal={IEEE transactions on pattern analysis and machine intelligence},
  volume={40},
  number={4},
  pages={769--790},
  year={2017},
  publisher={IEEE}
}

@inproceedings{GQR_L2H,
  title={A general and efficient querying method for learning to hash},
  author={Li, Jinfeng and Yan, Xiao and Zhang, Jian and Xu, An and Cheng, James and Liu, Jie and Ng, Kelvin KW and Cheng, Ti-chung},
  booktitle={Proceedings of the 2018 International Conference on Management of Data},
  pages={1333--1347},
  year={2018}
}

@inproceedings{IVF,
  title={Video Google: A text retrieval approach to object matching in videos},
  author={Sivic and Zisserman},
  booktitle={Proceedings ninth IEEE international conference on computer vision},
  pages={1470--1477},
  year={2003},
  organization={IEEE}
}

@article{IMI,
  title={The inverted multi-index},
  author={Babenko, Artem and Lempitsky, Victor},
  journal={IEEE transactions on pattern analysis and machine intelligence},
  volume={37},
  number={6},
  pages={1247--1260},
  year={2014},
  publisher={IEEE}
}

@inproceedings{FANNG,
  title={Fanng: Fast approximate nearest neighbour graphs},
  author={Harwood, Ben and Drummond, Tom},
  booktitle={Proceedings of the IEEE Conference on Computer Vision and Pattern Recognition},
  pages={5713--5722},
  year={2016}
}

@article{HNSW,
  title={Efficient and robust approximate nearest neighbor search using hierarchical navigable small world graphs},
  author={Malkov, Yu A and Yashunin, Dmitry A},
  journal={IEEE transactions on pattern analysis and machine intelligence},
  volume={42},
  number={4},
  pages={824--836},
  year={2020},
  publisher={IEEE}
}

@article{RNG,
  title={The relative neighbourhood graph of a finite planar set},
  author={Toussaint, Godfried T},
  journal={Pattern recognition},
  volume={12},
  number={4},
  pages={261--268},
  year={1980},
  publisher={Elsevier}
}

@article{NSG,
  title={Fast Approximate Nearest Neighbor Search With The Navigating Spreading-out Graph},
  author={Fu, Cong and Xiang, Chao and Wang, Changxu and Cai, Deng},
  journal={Proceedings of the VLDB Endowment},
  volume={12},
  number={5},
  pages={461--474},
  year={2019},
  publisher={VLDB Endowment}
}

@article{tau_MNG,
  title={Efficient approximate nearest neighbor search in multi-dimensional databases},
  author={Peng, Yun and Choi, Byron and Chan, Tsz Nam and Yang, Jianye and Xu, Jianliang},
  journal={Proceedings of the ACM on Management of Data},
  volume={1},
  number={1},
  pages={1--27},
  year={2023},
  publisher={ACM New York, NY, USA}
}

@inproceedings{ScaNN,
  title={Accelerating large-scale inference with anisotropic vector quantization},
  author={Guo, Ruiqi and Sun, Philip and Lindgren, Erik and Geng, Quan and Simcha, David and Chern, Felix and Kumar, Sanjiv},
  booktitle={International Conference on Machine Learning},
  pages={3887--3896},
  year={2020},
  organization={PMLR}
}

@inproceedings{SOAR,
  title={SOAR: improved indexing for approximate nearest neighbor search},
  author={Sun, Philip and Simcha, David and Dopson, Dave and Guo, Ruiqi and Kumar, Sanjiv},
  booktitle={Proceedings of the 37th International Conference on Neural Information Processing Systems},
  pages={3189--3204},
  year={2023}
}

@inproceedings{Bliss,
  title={Bliss: A billion scale index using iterative re-partitioning},
  author={Gupta, Gaurav and Medini, Tharun and Shrivastava, Anshumali and Smola, Alexander J},
  booktitle={Proceedings of the 28th ACM SIGKDD Conference on Knowledge Discovery and Data Mining},
  pages={486--495},
  year={2022}
}

@article{PQfs,
  title={Cache locality is not enough: high-performance nearest neighbor search with product quantization fast scan},
  author={Andr{\'e}, Fabien and Kermarrec, Anne-Marie and Le Scouarnec, Nicolas},
  journal={Proceedings of the VLDB Endowment},
  volume={9},
  number={4},
  pages={288--299},
  year={2015},
  publisher={VLDB Endowment}
}

@article{QuickerADC,
  title={Quicker adc: Unlocking the hidden potential of product quantization with simd},
  author={Andr{\'e}, Fabien and Kermarrec, Anne-Marie and Le Scouarnec, Nicolas},
  journal={IEEE transactions on pattern analysis and machine intelligence},
  volume={43},
  number={5},
  pages={1666--1677},
  year={2019},
  publisher={IEEE}
}

@inproceedings{Bolt,
  title={Bolt: Accelerated data mining with fast vector compression},
  author={Blalock, Davis W and Guttag, John V},
  booktitle={Proceedings of the 23rd ACM SIGKDD International Conference on Knowledge Discovery and Data Mining},
  pages={727--735},
  year={2017}
}

@article{ADSampling,
  title={High-dimensional approximate nearest neighbor search: with reliable and efficient distance comparison operations},
  author={Gao, Jianyang and Long, Cheng},
  journal={Proceedings of the ACM on Management of Data},
  volume={1},
  number={2},
  pages={1--27},
  year={2023},
  publisher={ACM New York, NY, USA}
}

@article{SPANN,
  title={SPANN: Highly-efficient Billion-scale Approximate Nearest Neighborhood Search},
  author={Chen, Qi and Zhao, Bing and Wang, Haidong and Li, Mingqin and Liu, Chuanjie and Li, Zengzhong and Yang, Mao and Wang, Jingdong},
  journal={Advances in Neural Information Processing Systems},
  volume={34},
  pages={5199--5212},
  year={2021}
}

@inproceedings{SPFresh,
  title={SPFresh: Incremental In-Place Update for Billion-Scale Vector Search},
  author={Xu, Yuming and Liang, Hengyu and Li, Jin and Xu, Shuotao and Chen, Qi and Zhang, Qianxi and Li, Cheng and Yang, Ziyue and Yang, Fan and Yang, Yuqing and others},
  booktitle={Proceedings of the 29th Symposium on Operating Systems Principles},
  pages={545--561},
  year={2023}
}

@article{FaissGPU,
  title={Billion-scale similarity search with {GPUs}},
  author={Johnson, Jeff and Douze, Matthijs and J{\'e}gou, Herv{\'e}},
  journal={IEEE Transactions on Big Data},
  volume={7},
  number={3},
  pages={535--547},
  year={2019},
  publisher={IEEE}
}

@inproceedings{DiskANN,
  title={DiskANN: fast accurate billion-point nearest neighbor search on a single node},
  author={Subramanya, Suhas Jayaram and Devvrit and Kadekodi, Rohan and Krishaswamy, Ravishankar and Simhadri, Harsha Vardhan},
  booktitle={Proceedings of the 33rd International Conference on Neural Information Processing Systems},
  pages={13766--13776},
  year={2019}
}

@inproceedings{HM-ANN,
  title={HM-ANN: efficient billion-point nearest neighbor search on heterogeneous memory},
  author={Ren, Jie and Zhang, Minjia and Li, Dong},
  booktitle={Proceedings of the 34th International Conference on Neural Information Processing Systems},
  volume={33},
  pages={10672--10684},
  year={2020}
}

@inproceedings{Song,
  title={Song: Approximate nearest neighbor search on gpu},
  author={Zhao, Weijie and Tan, Shulong and Li, Ping},
  booktitle={2020 IEEE 36th International Conference on Data Engineering (ICDE)},
  pages={1033--1044},
  year={2020},
  organization={IEEE}
}

@inproceedings{FPGA_ann_cvpr,
  title={Efficient Large-Scale Approximate Nearest Neighbor Search on OpenCL FPGA},
  author={Zhang, Jialiang and Li, Jing and Khoram, Soroosh},
  booktitle={2018 IEEE/CVF Conference on Computer Vision and Pattern Recognition},
  pages={4924--4932},
  year={2018},
  organization={IEEE}
}

@inproceedings{danopoulos2019fpga,
  title={Fpga acceleration of approximate knn indexing on high-dimensional vectors},
  author={Danopoulos, Dimitrios and Kachris, Christoforos and Soudris, Dimitrios},
  booktitle={2019 14th International Symposium on Reconfigurable Communication-centric Systems-on-Chip (ReCoSoC)},
  pages={59--65},
  year={2019},
  organization={IEEE}
}

@inproceedings{PASE,
  title={Pase: Postgresql ultra-high-dimensional approximate nearest neighbor search extension},
  author={Yang, Wen and Li, Tao and Fang, Gai and Wei, Hong},
  booktitle={Proceedings of the 2020 ACM SIGMOD international conference on management of data},
  pages={2241--2253},
  year={2020}
}

@article{AnalyticDB-V,
  title={AnalyticDB-V: a hybrid analytical engine towards query fusion for structured and unstructured data},
  author={Wei, Chuangxian and Wu, Bin and Wang, Sheng and Lou, Renjie and Zhan, Chaoqun and Li, Feifei and Cai, Yuanzhe},
  journal={Proceedings of the VLDB Endowment},
  volume={13},
  number={12},
  pages={3152--3165},
  year={2020},
  publisher={VLDB Endowment}
}

@inproceedings{Milvus,
  title={Milvus: A purpose-built vector data management system},
  author={Wang, Jianguo and Yi, Xiaomeng and Guo, Rentong and Jin, Hai and Xu, Peng and Li, Shengjun and Wang, Xiangyu and Guo, Xiangzhou and Li, Chengming and Xu, Xiaohai and others},
  booktitle={Proceedings of the 2021 International Conference on Management of Data},
  pages={2614--2627},
  year={2021}
}

@article{Milvus_v2,
  title={Manu: a cloud native vector database management system},
  author={Guo, Rentong and Luan, Xiaofan and Xiang, Long and Yan, Xiao and Yi, Xiaomeng and Luo, Jigao and Cheng, Qianya and Xu, Weizhi and Luo, Jiarui and Liu, Frank and others},
  journal={Proceedings of the VLDB Endowment},
  volume={15},
  number={12},
  pages={3548--3561},
  year={2022},
  publisher={VLDB Endowment}
}

@inproceedings{vearch,
  title={The design and implementation of a real time visual search system on JD E-commerce platform},
  author={Li, Jie and Liu, Haifeng and Gui, Chuanghua and Chen, Jianyu and Ni, Zhenyuan and Wang, Ning and Chen, Yuan},
  booktitle={Proceedings of the 19th International Middleware Conference Industry},
  pages={9--16},
  year={2018}
}

@inproceedings{VBASE,
  title={$\{$VBASE$\}$: Unifying Online Vector Similarity Search and Relational Queries via Relaxed Monotonicity},
  author={Zhang, Qianxi and Xu, Shuotao and Chen, Qi and Sui, Guoxin and Xie, Jiadong and Cai, Zhizhen and Chen, Yaoqi and He, Yinxuan and Yang, Yuqing and Yang, Fan and others},
  booktitle={17th USENIX Symposium on Operating Systems Design and Implementation (OSDI 23)},
  pages={377--395},
  year={2023}
}

@article{Starling,
  title={Starling: An I/O-Efficient Disk-Resident Graph Index Framework for High-Dimensional Vector Similarity Search on Data Segment},
  author={Wang, Mengzhao and Xu, Weizhi and Yi, Xiaomeng and Wu, Songlin and Peng, Zhangyang and Ke, Xiangyu and Gao, Yunjun and Xu, Xiaoliang and Guo, Rentong and Xie, Charles},
  journal={Proceedings of the ACM on Management of Data},
  volume={2},
  number={1},
  pages={1--27},
  year={2024},
  publisher={ACM New York, NY, USA}
}

@article{ANN_Benchmarks,
  title={ANN-Benchmarks: A benchmarking tool for approximate nearest neighbor algorithms},
  author={Aum{\"u}ller, Martin and Bernhardsson, Erik and Faithfull, Alexander},
  journal={Information Systems},
  volume={87},
  pages={101374},
  year={2020},
  publisher={Elsevier}
}

@online{Sift_Gist,
  author={Jegou, Herve and Amsaleg, Laurent},
  title={Datasets for approximate nearest neighbor search},
  year={2010},
  url={http://corpus-texmex.irisa.fr},
  urldate={2024-11-27}
}

@online{MSong,
  author={Lidy, Thomas},
  title={Million Song Dataset Benchmarks},
  year={2015},
  url={https://www.ifs.tuwien.ac.at/mir/msd},
  urldate={2024-11-27}
}

@online{T2I,
  author={Baranchuk, Dmitry and Babenko, Artem},
  title={Text-to-Image dataset for billion-scale similarity search},
  year={2021},
  url={https://research.yandex.com/datasets/text-to-image-dataset-for-billion-scale-similarity-search},
  urldate={2024-11-27}
}

@online{OpenAI5M,
  author={Zilliztech},
  title={VectorDBBench: A Benchmark Tool for VectorDB},
  year={2023},
  url={https://github.com/zilliztech/VectorDBBench},
  urldate={2024-11-27}
}

@article{Faiss,
  title={The Faiss library},
  author={Douze, Matthijs and Guzhva, Alexandr and Deng, Chengqi and Johnson, Jeff and Szilvasy, Gergely and Mazaré, Pierre-Emmanuel and Lomeli, Maria and Hosseini, Lucas and Jégou, Herve},
  year={2024},
  eprint={2401.08281},
  archivePrefix={arXiv},
  primaryClass={cs.LG}
}

@online{recommend_nlist,
  author={Douze, Matthijs},
  title={Guidelines to choose an index},
  year={2024},
  url={https://github.com/facebookresearch/faiss/wiki/Guidelines-to-choose-an-index},
  urldate={2024-9-6}
}

@incollection{col_filter_recommend,
  title={Collaborative filtering recommender systems},
  author={Schafer, J Ben and Frankowski, Dan and Herlocker, Jon and Sen, Shilad},
  booktitle={The adaptive web: methods and strategies of web personalization},
  pages={291--324},
  year={2007},
  publisher={Springer}
}

@article{nn_data_mining,
  title={Nearest neighbor pattern classification},
  author={Cover, Thomas and Hart, Peter},
  journal={IEEE transactions on information theory},
  volume={13},
  number={1},
  pages={21--27},
  year={1967},
  publisher={IEEE}
}

@article{liu2007survey,
  title={A survey of content-based image retrieval with high-level semantics},
  author={Liu, Ying and Zhang, Dengsheng and Lu, Guojun and Ma, Wei-Ying},
  journal={Pattern recognition},
  volume={40},
  number={1},
  pages={262--282},
  year={2007},
  publisher={Elsevier}
}

@inproceedings{weber1998quantitative,
  title={A quantitative analysis and performance study for similarity-search methods in high-dimensional spaces},
  author={Weber, Roger and Schek, Hans-J{\"o}rg and Blott, Stephen},
  booktitle={VLDB},
  volume={98},
  pages={194--205},
  year={1998}
}

@inproceedings{indyk1998approximate,
  title={Approximate nearest neighbors: towards removing the curse of dimensionality},
  author={Indyk, Piotr and Motwani, Rajeev},
  booktitle={Proceedings of the thirtieth annual ACM symposium on Theory of computing},
  pages={604--613},
  year={1998}
}

\clearpage

\appendix

\section{Proof of Theorem~\ref{theo:rair}}
\label{sec:proof}

\begin{proof}
\label{proof:rair}

Let $Q_l$ be the set of queries that are uniformly distributed over the hypersphere centered at $x$ with radius $l$.

We may expand the expectation as follows:
\[
\begin{aligned}
  &\quad L(c', c, Q_l) = E_{||R||=l}[{ReLU}(-\cos\alpha) \cdot (||q-c'||^2-||q-x||^2)] \\
  &= \int_{\frac{\pi}{2}}^{\pi} -\cos\alpha\ E_{||R||=l}[(R-r')^2-R^2\ |\ \frac{\langle R,r \rangle}{||R||\cdot||r||} = \alpha]\\
  &\ \ dP(\frac{\langle R,r \rangle}{||R||\cdot||r||} < \alpha)
\end{aligned}
\]

To evaluate this integral, we decompose $R$ and $r'$ into components parallel and orthogonal to $r$. 
As shown in Figure~\ref{fig:RAIR_geo}, we call these components $R_{\parallel}$, $R_{\perp}$, $r_{\parallel}'$, $r_{\perp}'$, respectively.
Also, we let $\theta=\angle cxc'$, $\beta$ is the angle between $R_{\perp}$ and $r_{\perp}'$.
Then we simplify this inner expectation given fixed $r$, $r'$, $\theta$, and $\alpha$:

\[
\begin{aligned}
&\quad E_{||R||=l}[(R-r')^2-R^2\ |\ \frac{\langle R,r \rangle}{||R||\cdot||r||} = \alpha] \\
&= E_{||R||=l}[(R_{\parallel}-r_{\parallel}')^2 + (R_{\perp}-r_{\perp}')^2 - R_{\parallel}^2 - R_{\perp}^2\ |\ ||R_{\parallel}|| = l\cos\alpha] \\
&= E_{||R||=l}[r_{\parallel}'(r_{\parallel}'-2R_{\parallel}) + r_{\perp}'^2-2r_{\perp}'R_{\perp}\ |\ ||R_{\parallel}|| = l\cos\alpha] \\
&= E_{||R||=l}[||r'|| \cos \theta (||r'|| \cos\theta - 2l\cos\alpha) + ||r'||^2 \sin^2 \theta - 2r_{\perp}'R_{\perp}\ |\\ 
&\quad ||R_{\parallel}|| = l\cos\alpha] \\
&= E_{\beta}[||r'||^2 - 2l||r'|| \cos\theta \cos\alpha - 2l||r'|| \sin\theta \sin\alpha \cos\beta] \\
&= ||r'|| \cdot (||r'|| - 2l\cos\theta \cos\alpha - 2l \sin\theta \sin\alpha E[\cos\beta]) \\
&= ||r'|| \cdot (||r'|| - 2l\cos\theta \cos\alpha)
\end{aligned}
\]

Meanwhile, $dP(\frac{\langle R,r \rangle}{||R||\cdot||r||} < \alpha)$ is proportional to the surface area of a $(D-1)$-dimensional hypersphere of radius $l \sin\alpha$.
Thus, we express it as $A \sin^{D-2} \alpha$ for some constant $A$.
Our integral then becomes

\[
\begin{aligned}
L(c', c, Q_l) &= \int_{\frac{\pi}{2}}^{\pi} -\cos\alpha ||r'|| (||r'|| - 2l\cos\theta \cos\alpha) A \sin^{D-2} \alpha d\alpha \\
&= - A||r'||^2 \int_{\frac{\pi}{2}}^{\pi} \cos\alpha \sin^{D-2} \alpha d\alpha \\
&\quad +\ 2Al||r'||\cos\theta \int_{\frac{\pi}{2}}^{\pi} \cos^2 \alpha \sin^{D-2} \alpha d\alpha \\
&= \frac{A||r'||^2}{D-1}\ +\ Al||r'||\cos\theta \int_{0}^{\pi} (\sin^{D-2} \alpha - \sin^{D} \alpha) d\alpha
\end{aligned}
\]

Now we define $I_D = \int_{0}^{\pi} \sin^D \alpha d\alpha$. 

\[
\begin{aligned}
I_D &= 2\int_{0}^{\pi/2} \sin^D \alpha d\alpha \\
&= -2 \cos \alpha \sin^{D-1} \alpha\ \bigg|_{0}^{\pi/2} \\
&\quad -2 \int_{0}^{\pi/2} \cos \alpha (-(D-1) \cos \alpha \sin^{D-2} \alpha) d\alpha \\
&= 2(D-1) \int_{0}^{\pi/2} \cos^2 \alpha \sin^{D-2} \alpha d\alpha \\
&= (D-1)I_{D-2} - (D-1)I_D
\end{aligned}
\]

Combining terms, we have $I_D = \frac {D-1}{D} I_{D-2}$. 
Then the loss satisfies:

\[
\begin{aligned}
L(c', c, Q_l) &= \frac{A||r'||^2}{D-1}\ +\ Al||r'||\cos\theta (I_{D-2} - I_D) \\
&= \frac{A||r'||^2}{D-1}\ +\ Al||r'||\cos\theta \frac{I_D}{D-1} \\
&= \frac{A}{D-1} (||r'||^2 + I_Dl \cos\theta ||r'||) \\
&= \frac{A}{D-1} (||r'||^2 + \frac{I_Dl}{||r||} r^T r') \\
\end{aligned}
\]

Note that the hypersphere $Q$ consists of many hyperspherical surface $Q_l$. We solve $L(c',c,Q)$:

\[
\begin{aligned}
L(c', c, Q) &= \int_{0}^{l_m} L(c', c, Q_l) \cdot \frac{Dl^{D-1}}{l_m^D} dl \\
&= \int_{0}^{l_m} \frac{A}{D-1} (||r'||^2 + \frac{I_Dl}{||r||} r^T r') \frac{Dl^{D-1}}{l_m^D} dl \\
&= \frac{A}{D-1}(||r'||^2 + \int_{0}^{l_m} \frac{D I_D l^{D}}{l_m^D||r||} r^T r' dl) \\
&= \frac{A}{D-1}(||r'||^2 + \frac{DI_Dl_m}{(D+1)||r||} r^T r') \\
&\propto ||r'||^2 + \lambda r^T r'
\end{aligned}
\]

where $\lambda = \frac{DI_Dl_m}{(D+1)||r||} > 0$.

\end{proof}

\begin{figure}[t]
  \centering
  \includegraphics[width=0.8 \linewidth]{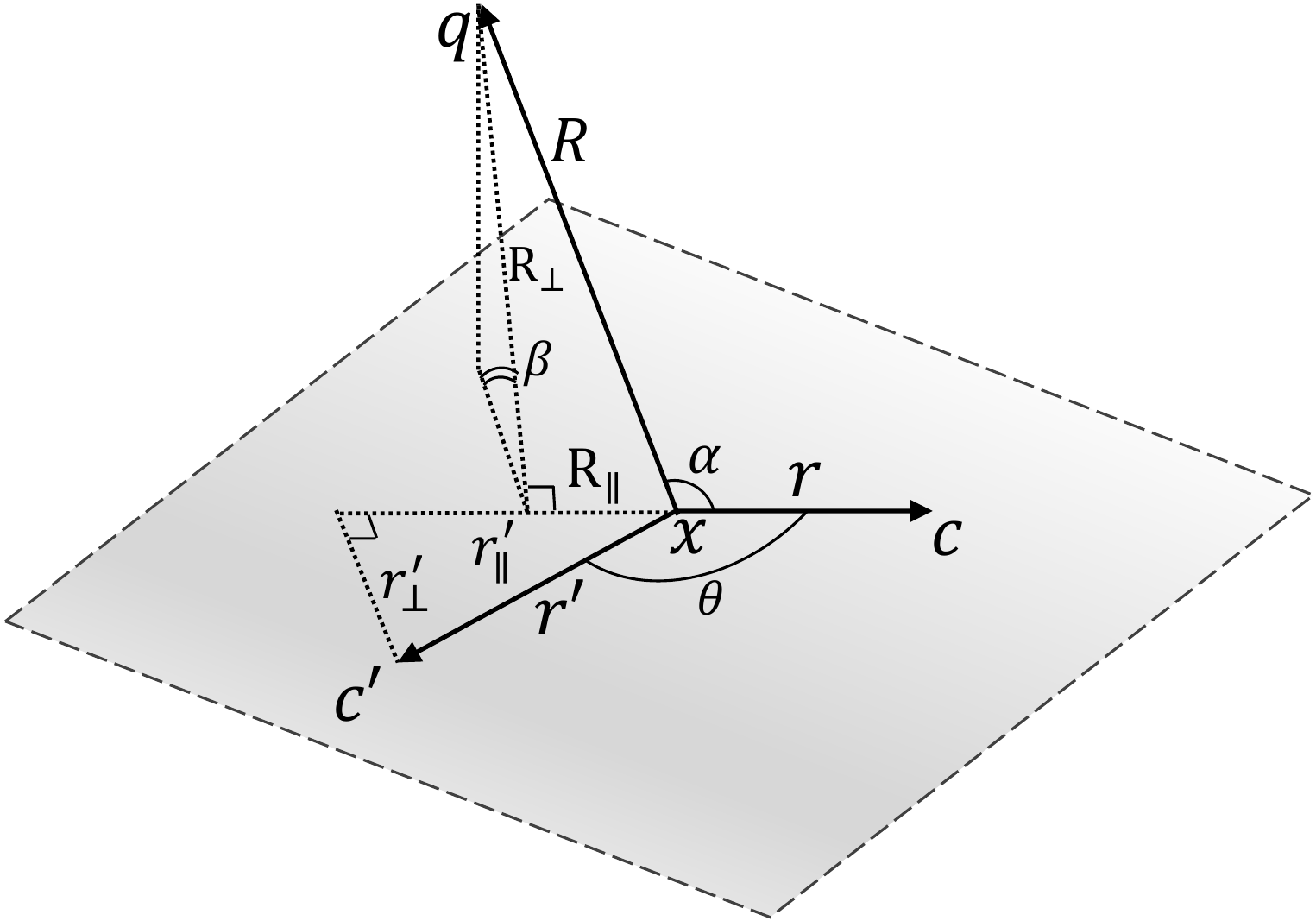}
  \caption{Detailed geometric relationship of vectors.}
  \label{fig:proof_geo}
\end{figure}

\end{document}